\documentclass{article}

\usepackage{PRIMEarxiv}

\usepackage[utf8]{inputenc} 
\usepackage[T1]{fontenc}    
\usepackage{hyperref}       
\usepackage{url}            
\usepackage{booktabs}       
\usepackage{amsfonts}       
\usepackage{nicefrac}       
\usepackage{microtype}      
\usepackage{fancyhdr}       
\usepackage{graphicx}       

\usepackage{float}
\usepackage{multirow}
\usepackage{paralist}
\usepackage{arydshln}
\usepackage{stfloats}
\usepackage{xspace}
\usepackage{braket}
\usepackage[most]{tcolorbox}
\usepackage{caption}
\usepackage{lscape}
\usepackage{ragged2e}
\usepackage[backend=biber,style=authoryear]{biblatex}
\usepackage{multirow}
\usepackage{makecell}

\newcommand{\survivalRate}{SR\xspace}
\hypersetup{
colorlinks = true, 
linkcolor=black, 
citecolor=black, 
urlcolor=black 
}

\title{Quantum Circuit Mutants: Empirical Analysis and Recommendations}

\author{
  Eñaut Mendiluze Usandizaga \\
  Simula Research Laboratory and\\
  Oslo Metropolitan University \\
  Oslo, Norway \\
  \texttt{enaut@simula.no} \\
   \And
  Tao Yue \\
  Simula Research Laboratory \\
  Oslo, Norway \\
  \texttt{taoyue@gmail.com} \\
   \And
   Paolo Arcaini \\
  National Institute of Informatics \\
  Tokyo, Japan \\
  \texttt{arcaini@nii.ac.jp} \\
   \And
   Shaukat Ali \\
  Simula Research Laboratory and\\
  Oslo Metropolitan University \\
  Oslo, Norway \\
  \texttt{shaukat@simula.no} \\
}

\begin{document}
\maketitle

\begin{abstract}
As a new research area, quantum software testing lacks systematic testing benchmarks to assess testing techniques' effectiveness. Recently, some open-source benchmarks and mutation analysis tools have emerged. However, there is insufficient evidence on how various quantum circuit characteristics (e.g., circuit depth, number of quantum gates), algorithms (e.g., Quantum Approximate Optimization Algorithm), and mutation characteristics (e.g., mutation operators) affect the detection of mutants in quantum circuits. Studying such relations is important to systematically design faulty benchmarks with varied attributes (e.g., the difficulty in detecting a seeded fault) to facilitate assessing the cost-effectiveness of quantum software testing techniques efficiently. To this end, we present a large-scale empirical evaluation with more than 700K faulty benchmarks (quantum circuits) generated by mutating 382 real-world quantum circuits. Based on the results, we provide valuable insights for researchers to define systematic quantum mutation analysis techniques. We also provide a tool to recommend mutants to users based on chosen characteristics (e.g., a quantum algorithm type) and the required difficulty of detecting mutants. Finally, we also provide faulty benchmarks that can already be used to assess the cost-effectiveness of quantum software testing techniques. 
\end{abstract}



\maketitle

\section{Introduction}

Quantum Computing (QC) is a fairly recent field that is advancing quickly~\parencite{evolutionQC}, promising to revolutionize computing by offering solutions to some complex problems with the enormous computational power of quantum computers. Quantum software empowers the QC application development~\parencite{ourCACM}. Naturally, there is a growing need to test quantum software to assess quantum software's correctness. To this end, several quantum software testing techniques have emerged in the last few years~\parencite{ourICST2021,honarvar2020property,QmutPy,mutation-based,QMutPy2,QMutPy3,genTestsQPSSBSE2021,CTQuantumQRS2021,quraTestASE23,pauliStringsASE2024}.

Quantum software testing techniques need benchmarks to assess their cost-effectiveness. To this end, some open-source benchmarks appeared recently~\parencite{zhao2021bugs4q,Bugs4QJSS2023,9474565,9814922}. However, such benchmarks are small-scale and do not provide systematic classifications of bug features that could be used to systematically assess the effectiveness of quantum software testing techniques. At the same time, some quantum mutation analysis techniques with tools have been published recently~\parencite{Mendiluze2021,QmutPy,QMutPy2,QMutPy3}. However, these tools generate too many mutants, which become infeasible to execute due to scarce QC resources. Even if the execution is not an issue, many mutants generated by these tools are redundant and are often too easy to detect; thereby, they are not useful for testing. Finally, these tools do not provide a systematic and intelligent way to generate a small subset of mutants with varied characteristics such that quantum software testing techniques can be assessed more systematically.

In general, there is no sufficient understanding of quantum mutations, e.g., \emph{which mutants} are difficult to detect, \emph{where to seed} a fault in a quantum circuit so that it is difficult to detect, and \emph{which types} of mutations are related to each algorithm type. To build such understanding, to generate new knowledge about quantum mutants, and to generate faulty benchmarks of different characteristics to systematically and efficiently assess the cost-effectiveness of quantum software testing techniques, we present results of a large-scale empirical evaluation with more than 700K faulty benchmarks generated with an existing quantum mutation analysis tool~\parencite{Mendiluze2021}. Each generated benchmark is a faulty version of an original quantum circuit and is called \emph{faulty benchmark}. 

This empirical evaluation aims to study various mutation characteristics (e.g., mutation operator types, gate types, position), quantum algorithms and their classification (e.g., Variational Quantum Eigensolver, Amplitude Estimation, Grover's algorithm), and circuit characteristics (e.g., circuit depth, the number of gates, number of entangled qubits) on the ``survivability'' of faulty benchmarks, i.e., whether the fault seeded in a benchmark can survive the fault detection of a quantum software testing technique. Our motivation for choosing survivability is that we want to assess the difficulty of detecting a seeded fault. To this end, such survivability indicates how various mutation and circuit characteristics, and quantum algorithms and their classification, play a role in the effectiveness of detecting the fault in a faulty benchmark with a given quantum software testing technique.

Based on the results, key observations are: First, we found that faulty benchmarks generated with the \textit{Add} mutation operator (i.e., adding a new quantum gate) have higher survivability than \textit{Remove} or \textit{Replace}. Second, regarding the position where a mutation operator is applied to create a faulty benchmark, applying the mutation at the beginning or end of the circuit results in higher survivability, suggesting that these mutations have a minor impact on the circuit's behavior. In contrast, mutations in the middle of the circuit are more likely to significantly alter its behavior, making them easier to detect.
Third, survivability is strongly related to the algorithm used. Notably, the algorithms that are designed to produce one dominant output, i.e., output with the highest probability (e.g., optimization algorithms), are likely to lead to high survival rates. Finally, we also found no significant correlation between the circuit complexity characteristics (e.g., the number of qubits in a circuit) and survivability.

Our main contribution is a comprehensive empirical study to generate new knowledge on understanding relationships of mutation characteristics, circuit characteristics, quantum algorithms, and their interactions with the survivability of faulty benchmarks. Additionally, based on the results of the empirical study, we provide a command line-based recommendation tool to assist users in generating a desired number of faulty benchmarks of varied survivabilities by considering characteristics of interested quantum algorithms. Finally, we provide a large-scale faulty benchmark set consisting of more than 700K faulty benchmark circuits that can be readily used to assess the cost-effectiveness of quantum software testing techniques.
%
%

\textit{Paper structure:} Section~\ref{sec:background} and Section~\ref{sec:relatedwork} present the background and related work, respectively. We present the design of the empirical study in Section~\ref{sec:design}, and results and discussion in Section~\ref{sec:results}. We provide a more detailed discussion in Section~\ref{sec:discussion}. We conclude the paper in Section~\ref{sec:conclusionAndRelated}.

\section{Background} \label{sec:background}
This section provides an overview of basic quantum computing concepts in Section~\ref{subsec:BackgroundQC}. We compare classical and quantum software testing in Section~\ref{subsec:BackgroundSoftwareTestingC&Q}, and explore mutation analysis in both classical and quantum contexts, Sections~\ref{subsec:BackgroundMutation} and~\ref{subsec:BackgroundQuantumMutation} respectively.

\subsection{Quantum Computing} \label{subsec:BackgroundQC}
The main difference between a quantum computer and a classical computer is the smallest data unit on which they perform computations. Such data unit in classical computing is a bit, whereas, in Quantum Computing (QC), it is a quantum bit (qubit). A classical bit can only have a value of 0 or 1 at a given time point, whereas a qubit can be in a \textit{superposition} state of 0 and 1. Superposition is one of the special quantum characteristics leading to quantum speedup. Another key quantum characteristic is \textit{entanglement}, where two or more qubits are connected to each other, i.e., they will always be in the same state~\parencite{yanofsky2008quantum}. 
Another special characteristic of QC is that it obeys the \textit{no-cloning theorem}~\parencite{Wootters1982}, i.e., one cannot simply copy or measure a quantum state since it will result in collapsing a quantum state from a superposition state to a definite state (classical state)~\parencite{wiseman2009quantum}. In QC, the collapse of a qubit is used to obtain the measurement result of a quantum operation~\parencite{wiseman2009quantum}.

\begin{figure*}[!tb]
\centering
\includegraphics[width=0.9\linewidth]{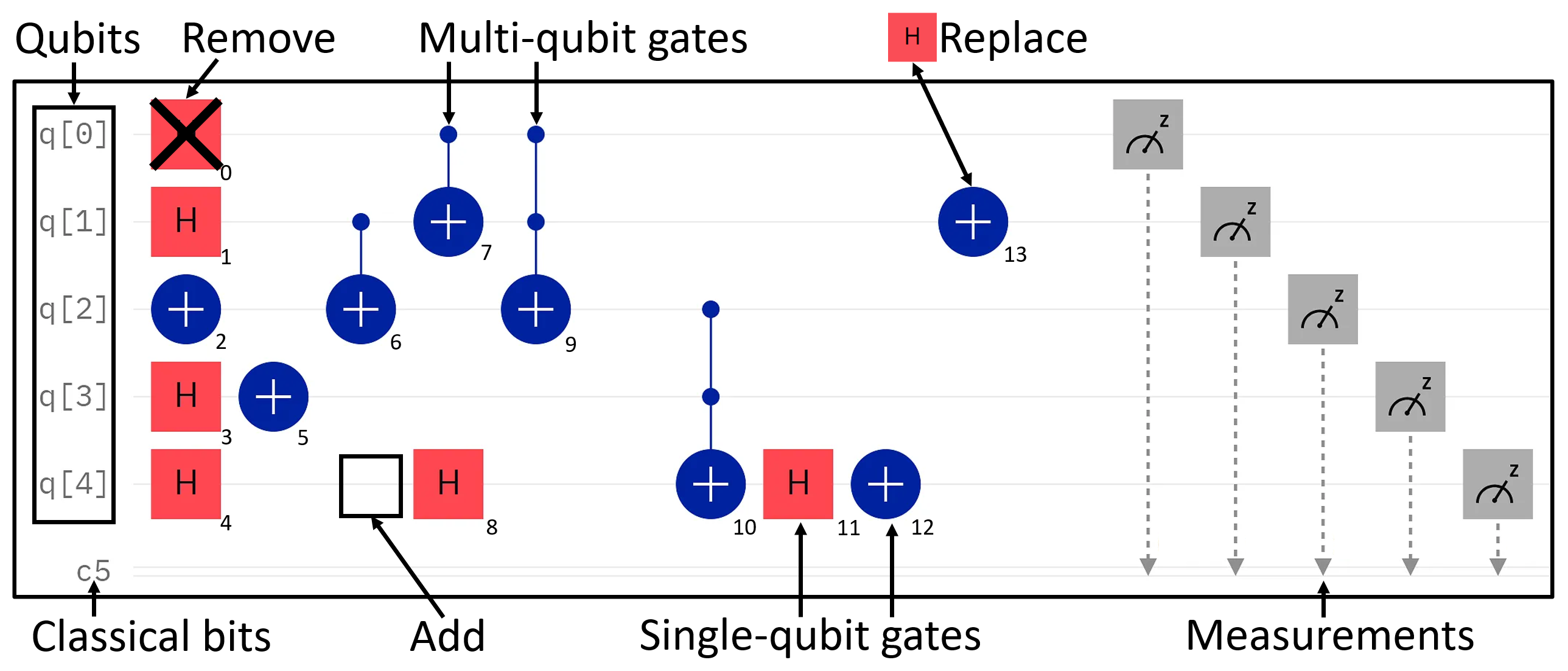}
\caption{A quantum circuit example.
The circuit has five qubits (i.e., $q[0]$ to $q[4]$) and a group of five classical bits (i.e., together denoted as $c_{5}$). 
The measurements collapse the state of all qubits to classical bits. The three selected operators are: \textit{Add} a gate before position 8 in the circuit, \textit{Remove} the Hadamard gate at position 0, and \textit{Replace} the NOT gate with a Hadamard gate at position 13.
}
\label{fig:QuantumCircuit}
\end{figure*}

Figure~\ref{fig:QuantumCircuit} shows an example of a quantum circuit visually drawn in IBM's quantum circuit composer~\parencite{IBMQuantumComposer}.
It shows the key elements of the circuit having five qubits ($q[0]$ to $q[4]$) and a group of five classical bits, i.e., $c_{5}$. A quantum circuit performs computations with quantum gates on qubits. A quantum gate takes qubit(s) as inputs and performs computation to alter the state of the quantum circuit. If a quantum gate operates on one qubit, we call it a \textit{Single Qubit Gate}, e.g., a Hadamard gate (\textit{H}). The Hadamard gate is represented as a square with an \textit{H} in Figure~\ref{fig:QuantumCircuit}, and is the gate used to put a qubit in a superposition state. A \textit{Multi-qubit Gate} operates on more than one qubit (e.g., conditional NOT gate (CNOT), where the control qubit is symbolized with a small filled circle, and the target qubit symbolized with a + sign within a circle). The conditional gates apply the operation of the gate only when the control qubit state is 1. In the case of the CNOT gate, the NOT operation is applied only if the conditional qubit's state is 1, flipping the state of the target qubit. At the end of the circuit, the measurements are shown. A measurement collapses the state of the qubit into a classical definite state, i.e., the quantum state is translated into a classical state and cannot be translated back to the quantum state.

\subsection{Classical Software Testing VS Quantum Software Testing} \label{subsec:BackgroundSoftwareTestingC&Q}
In classical software engineering, software testing has long been established as a fundamental aspect of ensuring the quality and reliability of software systems. Over the years, diverse testing techniques have been developed, ranging from basic unit testing to more sophisticated approaches such as model-based testing and regression testing~\parencite{IntroSWtesting}. Through systematic testing, software testers can detect and correct defects, enhancing their software system's functionality and robustness.

Classical software testing techniques often rely on methods such as observing and reproducing the software's execution state to check program behavior. Quantum software, operating under particular properties, such as quantum superposition, and entanglement, presents unique challenges when developing testing techniques. Unlike classical software, quantum states cannot be copied, making it difficult to directly observe and replicate internal states during execution. This restriction demands developing new testing approaches. Testing techniques must adapt to these properties, ensuring effectiveness even without being able to directly observe and analyze quantum states during execution~\parencite{ourCACM,zhao2021LandscapesAndHorizons,QST_SOTA,qseRoadmapTOSEM2025,quantumTestingRoadmapTOSEM2025}.

\subsection{Mutation Analysis} \label{subsec:BackgroundMutation}
Mutation analysis is a widely used technique in software engineering to evaluate the quality of testing techniques. It is a dynamic technique that assesses the testing technique's effectiveness by introducing small changes to the program and checking the behavior of the modified programs when executed.

A typical mutation analysis process systematically uses mutation operators to introduce changes to the correct program. As a result, each change produces a faulty version (\emph{mutant}) of the program. Test cases are executed on mutants to determine which test cases can detect the mutants. The testing results measure the adequacy of testing techniques used to generate test cases~\parencite{MutationOrigin}.

The mutants can have different behaviors depending on the operator applied and the program itself. Within this context, each mutant can be categorized into several groups. First, regarding whether the testing technique can detect it or not, a mutant is classified as {\it detected} (or {\it killed}), or as a {\it survivor}, i.e., not detected. On the other hand, those mutants that are not even able to be executed, due to some syntactical errors, are called {\it stillborn} mutants~\parencite{MutationSurvey}.

Each group is further divided into some subcategories. Inside the not-detected mutants, one of the main challenges of mutation analysis is related to {\it equivalent mutants}. Equivalent mutants are those that behave the same as the original program for all the inputs, thus, they cannot be detected by any test.
Moreover, another critical category is constituted by {\it redundant mutants} that make minimal contributions to the testing process. Those are the ones that are detected whenever other mutants are detected. This category includes {\it duplicated} mutants, which are equivalent to each other but not to the original program, and {\it subsumed} mutants, jointly detected when other mutants are detected. Any error detected in a {\it subsumed} mutant is also identified by the first mutant~\parencite{36Survey,37Survey,38Survey,39Survey}.

Classically, each detected mutant is also categorized depending on its strength.
{\it Weakly detected mutants} that expose differences in the program state immediately after execution compared to the execution of the original program; {\it firmly detected mutants} that expose differences at a later point; and {\it strongly detected mutants} that show observable differences in the outputs~\parencite{IntroSWtesting}.

\subsection{Quantum Mutation Analysis} \label{subsec:BackgroundQuantumMutation}
In quantum mutation analysis, we introduce faults in quantum circuits by introducing small changes at the quantum gate level. For instance, those faults can be introduced by adding, removing, or replacing quantum gates in an original quantum circuit. These are referred to as \textit{mutation operator types} in~\parencite{Mendiluze2021}. Figure~\ref{fig:QuantumCircuit} shows some examples of these operator types: replacing a NOT gate (\textit{+}) with a Hadamard gate (\textit{H}), deleting a Hadamard gate (\textit{H}), and adding any gate in a given position. Such a way of creating faulty versions of quantum circuits for assessing testing techniques has been performed in some existing works of quantum software testing~\parencite{honarvar2020property,ourICST2021,quitoASE21tool,Mendiluze2021,QmutPy,QMutPy2,QMutPy3}. 

Another important characteristic is the \textit{gate type}, introduced in~\parencite{Mendiluze2021}. \textit{Gate type} refers to the quantum gate present in the change applied to the circuit, i.e., we can add a new quantum gate to a quantum circuit, depending on factors such as the available number of qubits, and the supported gate set, whereas deletion or replacement operations are limited to the quantum gates already present in the circuit.
The third characteristic is the {\it position}, which refers to a particular place in the quantum circuit where a fault can be introduced. Each of the gates in the circuit is located in a specific position. When applying an operator, a position of the circuit is chosen, and depending on the change that is wanted, the change is applied in the gate linked to that position (Remove or Replace) or before the gate related to that position (Add). The position of a gate is determined by the order of the statements in the code. The first position refers to the first gate applied to the circuit, whereas the last position of the circuit is the last gate before the measurements. In Figure~\ref{fig:QuantumCircuit} the position of each gate is indicated with a number, e.g., removing a Hadamard gate in position 0, adding a new gate in position 8, or replacing a NOT gate with a Hadamard gate in position 13. Further details can be consulted in~\parencite{Mendiluze2021}.

In quantum mutation analysis, we primarily focus on evaluating the effect of mutations on the output of quantum circuits. This approach is necessary because intermediate states cannot be inspected without cutting the circuit or collapsing the superposition state. As a result, the analysis is centered on assessing the impact of mutations that produce significant changes in the circuit's final output. By concentrating on these {\it strongly killed (detected) mutants}, as referred to in classical mutation testing~\parencite{MutationSurvey}, we ensure the analysis captures the most impactful alterations to the quantum state, while aligning with the constraints of quantum computation.

\section{Related Work} \label{sec:relatedwork}
In this section, we review prior research on mutation analysis, in classical and quantum approaches Sections~\ref{subsec:RelatedClassicMutation} and~\ref{subsec:RelatedQuantumMutation} respectively. We also explore quantum software bug repositories in Section~\ref{subsec:RelatedQuantumBug} and discuss advancements in quantum software testing in Section~\ref{subsec:RelatedQuantumSoftware}. Moreover, we checked formal verification techniques for quantum programs in Section~\ref{subsec:RelatedFormal} and the role of static analysis in quantum software development in Section~\ref{subsec:RelatedStatic}.

\subsection{Classical Mutation Analysis}\label{subsec:RelatedClassicMutation} 
The surveys of Jia and Harman~\parencite{MutationSurveyOld} and Papadakis et al.~\parencite{MutationSurvey} show the growing interest of the research community in mutation analysis. Inside mutation analysis, to improve and adapt the mutation to different use cases, several different strategies have been developed. Some of them involve new mutant generation techniques or new mutant selection strategies~\parencite{MutationSurveyOld,MutationSurvey}. Different works have proposed operators for specific programming languages~\parencite{survey77,survey80,survey79,survey81,survey82}, categories of programming languages~\parencite{survey83,survey84,survey85,survey86}, categories of applications~\parencite{survey87,survey88,survey89,survey91,survey93,survey95}, and specific bug categories~\parencite{survey95,survey104,survey107,survey109,survey111,regexTestGenSTVR2018}. Other works, instead, focused on devising new reduction strategies~\parencite{survey112,survey113,survey114,survey115,survey116,survey117}.

Papadakis et al.~\parencite{MutationSurvey} propose a checklist of best practices for using mutation testing in the context of controlled experiments for classical programs. Instead, we focus on quantum circuits and provide guidelines to select mutants for quantum circuits based on an extensive empirical evaluation. 

Different works assessed the effectiveness of mutation operators for different artifacts, as we do in this work for mutation operators for quantum programs.
Smith et al.~\parencite{smith2007empirical} conducted an empirical evaluation to assess the effectiveness of MuJava's mutation operators~\parencite{MaICSE2006} in software testing. The study categorized the behavior of mutants generated by selected mutation operators during successive attempts to detect them. The categorization used in the study includes the crossfire (subsumed mutants), dead on arrival (mutants detected by the initial test suite without any specific focus on mutation testing), Killed, and Equivalent mutants. This categorization provides a deeper understanding of the performance of individual operators and the behaviors exhibited by their resultant mutants. Our work studies the quantum mutation operators, and quantum mutants generated by them, across different quantum circuits. Our study presents a ranking for different mutants and their characteristics, providing valuable observations about their behavior and use.

Just et al.~\parencite{just2014mutants} conducted a study to assess whether mutants can be an alternative to real faults. The study uses real faults from subject programs and compares the effectiveness of developer-written and automatically generated test suites in detecting these faults. The study investigated the correlation between real faults and mutants generated by commonly used mutation operators. It found that a statistically significant correlation exists for 73\% of the real faults. The study also proposed improvements to the mutation analysis technique by introducing new or stronger mutation operators. They observed that 17\% of faults were not coupled to any mutants, which reveals a fundamental limitation of mutation analysis, and the other 10\% of actual faults required implementing new mutation operators. Even though our study does not directly relate to real faults, it is a substantial empirical study that can be used as a reference point for associating them with real quantum faults.

Zhang et al.~\parencite{zhang2016towards} explored how mutation analysis can be used for assessing the quality of use case models with use case specifications detailed in restricted natural language, with the ultimate goal of supporting requirements inspection. They proposed a taxonomy of defect types and defined nearly 200 mutation operators. A set of case studies demonstrated the feasibility of the proposed mutation analysis methodology. In contrast to this work, our study focuses on mutation analysis for quantum circuits.

\subsection{Quantum Mutation Analysis}\label{subsec:RelatedQuantumMutation} 
\label{subsec:quantummutationanalysis} Two key mutation analysis tools are available, Muskit~\parencite{Mendiluze2021} and QMutPy~\parencite{QmutPy,QMutPy2,QMutPy3}. Both tools can generate mutated quantum circuits with mutation operators, e.g., related to adding, removing, or replacing quantum gates. Note that QMutPy provides additional mutation operators related to the measurement gates, Quantum Measurement Insertion, and Quantum Measurement Deletion. QMutPy further looks into bug patterns identified in~\parencite{BugPattern} and mutating a quantum gate with another ``syntactically-equivalent'' with the same number and types of arguments. Fortunato et al. point out that such equivalence helps reduce the total number of mutants and decreases the possibility of having equivalent mutants. 

QMutPy and Muskit could generate many mutants, which might be infeasible to execute and even redundant. This paper instead studies the influence of various mutation characteristics, circuit characteristics, and algorithms and their classifications on the survivability of mutants via large-scale empirical evaluation to collect evidence to support researchers and practitioners in selecting meaningful mutants for mutation analysis. Moreover, based on the evidence collected, diverse mutant generation strategies can be developed and possibly integrated into QMutPy or Muskit. 


Wang et al.~\parencite{mutation-based} proposed a mutation-based approach (\textit{MutTG}) for generating the minimum number of test cases that maximizes the number of detected mutants to save the cost required to execute many test cases. \textit{MutTG} also defines a metric to measure the difficulty of detecting a mutant based on the number of inputs that can detect the mutant out of the total number of inputs. The higher the number of inputs detecting the mutant, the easier it is to detect the mutant. Instead, we study the relationships between different circuit and mutant characteristics and quantum algorithms on the survivability of mutants with a large-scale empirical evaluation.

\subsection{Quantum Software Bug Repositories}\label{subsec:RelatedQuantumBug} \label{subsec:bugrepo} Recently, a few bug repositories have been published. The Bugs4Q benchmark suite~\parencite{zhao2021bugs4q} collects bugs from the Qiskit GitHub repository, ensuring each bug has a buggy and a fixed version. They collected 36 bugs. In~\parencite{9474565}, a proposal is presented for reproducible bugs in quantum software with a set of quantum programs, their corresponding bugs, and infrastructure to support experimentation. In~\parencite{9814922}, a multi-lingual benchmark for property-based testing of quantum programs coded in Q\# is proposed, consisting of a set of programs in Q\# and corresponding properties. These works provide small-scale benchmarks and do not systematically analyze mutant characteristics, thereby giving no evidence about how easy it is to detect which mutants and which circuit characteristics make it difficult to detect a mutant. Thus, this paper presents a large-scale empirical evaluation to study how various mutant and circuit characteristics, algorithms, and their interactions affect detecting mutants.

\subsection{Quantum software testing}\label{subsec:RelatedQuantumSoftware} 
The survey by García de la Barrera et al.~\parencite{QST_SOTA} shows an overview of the latest advancements in quantum software testing by systematically mapping the literature to evaluate its current state. The survey demonstrates an increasing trend in recent years to adapt classical testing techniques and develop new quantum software testing techniques. The study identifies three main trends in quantum software testing: using statistical methods on repeated executions to handle quantum stochasticity, adapting the Hoare logic~\parencite{hoareLogic} for quantum software correctness, and using quantum circuit reversibility for information conservation. The study highlights that, despite the ongoing efforts to integrate test engineering practices into quantum computing, a need for established frameworks to integrate best practices and techniques remains.

Several quantum software testing techniques have been adapted from classical software testing. Moreover, new techniques specific to quantum software testing have been developed, contributing to the advancement of quantum software testing. For example, Honarvar et al.~\parencite{honarvar2020property} proposed a property-based testing approach that defines properties over input and output and uses these properties for test generation; Abreu et al.~\parencite{metamorphic} proposed a metamorphic testing approach in the context of quantum software; Wang et al.~\parencite{CTQuantumQRS2021}, instead, proposed a combinatorial testing approach specifically designed for quantum software and investigated its effectiveness in detecting faults in quantum circuits. In contrast, our study provides valuable insights into the characteristics of quantum software mutants, which can serve as a foundation for assessing the efficacy and reliability of new or existing quantum software testing techniques.


\subsection{Formal verification for quantum programs}\label{subsec:RelatedFormal} 
Formal verification for quantum programs involves applying formal methods to check the correctness of quantum circuits. This new field of study tries to address the challenges introduced by the unique properties of quantum computers regarding new types of errors and the development of complex quantum algorithms. Unlike the empirical assessment employed by software testing, formal verification aims to provide a mathematical proof of the correctness of software behavior. The survey by Lewis et al.~\parencite{lewis2023formal} provides some valuable insights about the current state of the field and tools developed, such as QWire~\parencite{rand2018qwire} and SQIR~\parencite{hietala2020proving}. QWire emerged as one of the initial quantum programming languages for developing verifiable programs, whereas SQIR is a proof assistant for writing proofs about quantum programs. The survey highlights the importance of verification techniques that follow quantum software development, showing that verifiable programming languages need to be developed to be easy to learn or use.

Formal quantum verification methods provide proofs for quantum programs but require extensive quantum mechanics knowledge. Mutation analysis can be used to assess formal verification techniques by evaluating their ability to detect and handle mutations in system models or specifications as done for classical formal verification by Rao et al.~\parencite{MutationVerification}, and Jain et al.~\parencite{mCoq}. Our study provides more understanding of actual quantum mutation operators, which could potentially be useful to assess the effectiveness of formal verification methods for quantum programs in the future.

\subsection{Static analysis for quantum programs}\label{subsec:RelatedStatic} 
Static analysis involves analyzing code without executing it and identifying potential faults through code inspection, bug patterns, and heuristics. Zhao et al.~\parencite{Zhao2023} presented QChecker, a static analysis tool specifically designed for quantum programs written in Qiskit. QChecker was evaluated using the Bugs4Q benchmark and demonstrated effectiveness in detecting various bug types, such as incorrect use of parameters or backend importing errors. Paltenghi and Pradel~\parencite{Paltenghi2024} developed LintQ, a static analysis framework tailored for quantum programs, specifically targeting Qiskit written programs. It introduces abstractions for quantum-specific concepts like circuits, gates, and qubits, enabling the detection of domain-specific bugs. LintQ achieved a high precision of 91\%, identifying numerous bugs overlooked by prior methods, mostly measurement-related bugs such as incorrect placement or redundant measurements.

While static analysis can provide insights into possible issues at the code level, it cannot capture behavior at runtime, which is essential for revealing faults that only appear during program execution. On the other hand, dynamic testing allows one to observe the runtime behavior of the program. Mutation analysis is used to assess the quality of tests for this type of testing.

\section{Methodology \& Experiment Design} \label{sec:design}
We first identify and explain the research questions in Section~\ref{subsec:RQs} followed by the definition of various characteristics in Section~\ref{subsec:characteristics}. We describe metrics, subject systems, mutant generation, and experimental setup together with execution in Sections~\ref{subsec:metrics}--\ref{subsec:experimentalsetupexecution}, respectively. Finally, in Section~\ref{subsec:threatsToValidity}, we discuss threats to validity.

\subsection{Research Questions} \label{subsec:RQs}
\begin{itemize}
\item \textbf{RQ1}: How do the various quantum mutation characteristics influence the survivability of faulty benchmarks? This RQ is further divided into three sub-research questions:
\begin{itemize}
\item \textit{\textbf{RQ1.1}} How does each of the characteristics individually affect the survivability of faulty benchmarks? 
\item \textit{\textbf{RQ1.2}} How does each of the pair-wise combinations between characteristics affect the survivability of faulty benchmarks?
\item \textit{\textbf{RQ1.3}} How do the interactions among all the characteristics affect the survivability of faulty benchmarks? 
\end{itemize}
In RQ1, we study the main mutation characteristics (Section~\ref{subsubsec:mutationcharacter}) and their interactions. 
\item \textbf{RQ2}: How does a quantum algorithm or quantum algorithm type affect the survivability of faulty benchmarks?
RQ2 studies individual algorithms and the effect of their two types of outputs (Section~\ref{subsubsec:algorithmcharac}) on the survivability of faulty benchmarks:
\begin{itemize}
\item \textit{\textbf{RQ2.1}} How does the \textit{output dominance} affect the survivability of faulty benchmarks?
\item \textit{\textbf{RQ2.2}} How does each \textit{algorithm group} affect the survivability of faulty benchmarks?
\item \textit{\textbf{RQ2.3}} How does each \textit{algorithm} affect the survivability of faulty benchmarks? 
\end{itemize}
\item 

\textbf{RQ3}: How do gate and circuit complexity affect the survivability of its faulty benchmarks? In particular, we study the influence of circuit and gate characteristics on survivability.

\item \textbf{RQ4}: How do the interactions of the algorithm characteristics with the mutation characteristics influence the survivability of faulty benchmarks? This RQ is further divided into three sub-research questions:
\begin{itemize}
\item \textit{\textbf{RQ4.1}} How does the interaction between the \textit{algorithm} and mutation characteristics influence the survivability of faulty benchmarks?
\item \textit{\textbf{RQ4.2}} How does the interaction between the \textit{algorithm group} and mutation characteristics influence the survivability of faulty benchmarks?
\item \textit{\textbf{RQ4.3}} How does the interaction between the \textit{output dominance} and mutation characteristics influence the survivability of faulty benchmarks?
\end{itemize}
\end{itemize}

Note that we also studied interactions with more than one mutation characteristic. However, given many possible combinations, they are only considered when automatically generating recommendations (Section~\ref{sec:discussion}). Nonetheless, all interaction data is available in our repository~\parencite{GitHubRepository}.

\subsection{Characteristics of Mutations, Circuits, and Algorithms -- Independent Variables} \label{subsec:characteristics}

\subsubsection{Mutation Characteristics}\label{subsubsec:mutationcharacter}

\noindent\textbf{Mutation Operator Type (\textit{Operator)}:} We have three types of mutation operators, i.e., adding (\textit{Add}), removing (\textit{Remove}), and replacing (\textit{Replace}) a quantum gate as described in Section~\ref{sec:background}.

\noindent\textbf{Quantum Gate Mutations:} We study quantum gate characteristics from three perspectives. First, we study \textit{mutated gates} (\textit{Gate}) such as \textit{Hadamard} and \textit{CNOT} with a mutation operator (e.g., \textit{Add}). In total, we have 19 gates that are currently implemented in Muskit~\parencite{Mendiluze2021}, which we used as a mutation framework. These gates are \textit{ccx}, \textit{cswap}, \textit{cx}, \textit{cz}, \textit{h}, \textit{id}, \textit{p}, \textit{rx}, \textit{rxx}, \textit{ry}, \textit{rz}, \textit{rzz}, \textit{s}, \textit{swap}, \textit{sx}, \textit{t}, \textit{x}, \textit{y}, and \textit{z}. Interested readers may refer to the following reference for more details about each gate~\parencite{weaver2022qiskit}. Second, we study \textit{mutated gate types} (\textit{Gate Type}) by classifying the implemented gates in Muskit 
into these seven categories representing the basic building blocks of Qiskit, i.e., Controlled gates (\textit{Controlled}), Hadamard gates (\textit{Hadamard}), Pauli gates (\textit{Pauli}), Phase gates (\textit{Phase}), Rotation gates (\textit{Rotation}), Swap gates (\textit{Swap}), and T gates (\textit{T}). Third, we study \textit{mutated gate size} (\textit{Gate Size}), which classifies quantum gates into two categories: single-qubit (\textit{Single}) and multi-qubit gates (\textit{Multi}). This classification is common in quantum circuit design~\parencite{weaver2022qiskit}. Note that these three independent variables (i.e., \textit{Gate}, \textit{Gate Type}, and \textit{Gate Size}) are intertwined, which makes it very difficult to interpret their interaction effects. As a result, we do not study their interactions.

\noindent\textbf{Position (\textit{Position}):} We study the position in the circuit where a change is introduced as described in Section~\ref{sec:background}. Given that the total number of positions varies from one circuit to another, we use the \textit{relative position} to the whole in terms of percentage, i.e., 10\%, 20\%, \ldots, 100\% to describe the position in the quantum circuit where a fault is seeded. For instance, 10\% means the first 10\% of the positions in a circuit.

\subsubsection{Algorithms Characteristics} \label{subsubsec:algorithmcharac}
\textbf{Algorithms:} We study the effect of various algorithms (\textit{Algorithm}) and their categorization from two aspects on the survivability of a faulty benchmark. We have 28 algorithms from MQT Bench, i.e., Quantum Program (QP) 1--28 as shown at the bottom of Table~\ref{tab:ProgramsCharacteristics}. More details about these algorithms can be found in~\parencite{quetschlich2023mqtbench}. Moreover, we use a classification from~\parencite{quetschlich2023mqtbench} to classify the 28 algorithms into 12 categories and name this classification as \textit{Algorithm Group}. These 12 categories are \textit{ae}, \textit{dj}, \textit{ghz}, \textit{graphstate}, \textit{grover}, \textit{qaoa}, \textit{qft}, \textit{qgan}, \textit{qpe}, \textit{qwalk}, \textit{vqe}, and \textit{wstate}. Such categories comprise the QPs that use the same \textit{Algorithm} as a building block, i.e., derived from the same \textit{Algorithm}. For example, in the case of the \textit{ae} \textit{Algorithm Group}, we consider the quantum programs based on \textit{ae}: QP1, QP12, and QP13. QP1 consists of the original \textit{ae} algorithm, and QP12 and QP13 are specific versions of the \textit{ae} algorithm to satisfy a specific problem. QP12 uses the \textit{ae} algorithm iteratively to estimate the fair price of a European call option, and QP13 uses the \textit{ae} algorithm to estimate the fair price of a European put option. Some of the \textit{Algorithms} are not used in other \textit{Algorithms} as building blocks, meaning that some of the \textit{Algorithm Group} consist only of one \textit{Algorithm}. A brief explanation of each \textit{Algorithm Group} together with the group \textit{Algorithms} is shown in Table~\ref{tab:algorithm_groups}. 
\begin{table}[!tb]
\centering
\small
\caption{Algorithm Groups details}
\label{tab:algorithm_groups}
\resizebox{\linewidth}{!}{
\begin{tabular}{p{0.15\linewidth}p{0.6\linewidth}c}
\toprule
\centering\textit{\textbf{Algorithm Group}} & \centering\textbf{Description} & \textit{\textbf{Algorithms}} \\
\midrule
\centering\textit{dj} & \centering Algorithms based on the Deutsch-Jozsa algorithm, which determines whether a given function is constant or balanced. & QP2 \\\\
\centering\textit{ghz} & \centering Algorithms based on the Greenberger-Horne-Zeilinger state, which represents a maximally entangled quantum state of multiple qubits where all qubits are simultaneously in a superposition state. & QP3 \\\\
\centering\textit{graphstate} & \centering Algorithms based on the Graph-state, which represents a type of entangled quantum state encoded based on the topology of a graph. & QP4 \\\\
\centering\textit{qft} & \centering Algorithms based on the Quantum Fourier Transform, the quantum equivalent of the discrete Fourier transform. & QP15, QP16 \\\\
\centering\textit{qgan} & \centering Algorithms based on the Quantum Generative Adversarial Network, a hybrid quantum-classic algorithm used for generative modeling tasks. & QP17 \\\\
\centering\textit{qwalk} & \centering Algorithms based on quantum walks, equivalent of classic random walks for the quantum paradigm. & QP20, QP21 \\\\
\centering\textit{wstate} & \centering Algorithms based on the W state, which involves entangling a single qubit with the collective state of the remaining qubits. & QP28 \\\\
\centering\textit{ae} & \centering Algorithms based on the Amplitude estimation algorithm, which finds an estimation of the amplitude of a certain quantum state. & QP1, QP12, QP13 \\\\
\centering\textit{grover} & \centering Algorithms based on Grover's algorithm, which finds a certain goal quantum state determined by an oracle. & QP8, QP9 \\\\
\centering\textit{qaoa} & \centering Algorithms based on the Quantum Approximation Optimization Algorithm, consisting of a parametrizable quantum algorithm to solve optimization problems. & QP10, QP14 \\\\
\centering\textit{qpe} & \centering Algorithms based on the Quantum Phase Estimation, which estimates the phase of a quantum state. & QP18, QP19 \\
\centering\textit{vqe} & \centering Algorithms based on the Variation Quantum Eigensolver, used to find the ground state of a given physical system. & \begin{tabular}[c]{@{}c@{}}\\QP5, QP6, QP7, QP11,\\ QP22, QP23, QP24,\\ QP25, QP26, QP27\end{tabular}\\ \bottomrule
\end{tabular}}
\\
\justifying
\tiny*QP1: Amplitude Estimation (ae); QP2: Deutsch-Jozsa (dj); QP3: Greenberger-Horne-Zeilinger State (ghz); QP4: Graph State (graphstate); QP5: Ground State (groundstatelarge); QP6: Ground State (groundstatemedium); QP7: Ground State (groundstatesmall); QP8: Grover Search without Ancilla (grover-noancilla); QP9: Grover Search with Ancilla (grover-v-chain); QP10: Portfolio Optimization with QAOA (portfolioqaoa); QP11: Porfolio Optimization with VQE (portfoliovqe); QP12: Pricing Call Option (pricingcall); QP13: Pricing Put Option (pricingput); QP14: Quantum Approximate Optimization Algorithm (qaoa); QP15: Quantum Fourier Transform (qft); QP16: Quantun Fourier Transform Entangled (qftentangled); QP17: Quantum Generative Adversarial Networks (qgan); QP18: Quantum Phase Estimation Exact (qpeexact); QP19: Quantum Phase Estimation Inexact (qpeinexact); QP20: Quantum Walk without Ancilla (qwalk-noancilla); QP21: Quantum Walk with Ancilla (qwalk-v-chain); QP22: Real Amplitudes ansatz with Random Parameters (realamprandom); QP23: Routing Algorithm (routing); QP24: Efficient SU2 ansatz with Random Parameters (su2random); QP25: Travelling Salesman (tsp); QP26: Two Local ansatz with random parameters (twolocalrandom); QP27: Variational Quantum Eigensolver (vqe); QP28: W-State (wstate).
\end{table}
An interested reader can consult~\parencite{quetschlich2023mqtbench} for more details on it.

In addition, we classify all the algorithms into two categories according to their \textit{Output Dominance}: (1) \textit{output-dominant} algorithms that focus on finding the output with the highest probability, such as the case for optimization algorithms. For such algorithms, we check if the produced dominant output matches the expected one. In total, we have 19 output-dominant algorithms; (2) \textit{diverse-output} algorithms with many outputs of different probabilities. As a result, to check the correctness of diverse-output algorithms, we need to compare all possible outputs and their probabilities with the expected ones. In total, we have nine of these algorithms.

\subsubsection{Circuit Characteristics} \label{subsubsec:circuitcharac}

\noindent\textbf{Circuit Complexity:} We study the typical metrics used to measure the complexity of circuits, i.e., the \textit{number of qubits} (\textit{\#qubits}), the \textit{total number of quantum gates} (\textit{\#gates}), and the \textit{number of measurements} (\textit{\#measurements}), counting the numbers of qubits, gates, and measurements in a circuit~\parencite{azad2021circuit,QASMBench}. In addition, we use circuit depth ($depth$), a commonly used metric, to measure the complexity of a quantum circuit, which is defined as the length of the longest path (measured as the number of gates) of the circuit from its beginning to the end~\parencite{childs_et_al}. 

\noindent \textbf{Gate Complexity:}
To assess the effect of gate complexity on survivability, we study three characteristics, i.e., the number of single-qubit gates (\textit{\#singleGates}), the number of multi-qubit gates (\textit{\#multiGates}), and the number of entangled qubits (\textit{\#eQubits}). We count the number of entangled qubits in a circuit by checking all its interaction states in the circuit and the qubits they relate to.

\subsection{Metrics and Statistical Tests} \label{subsec:metrics}

To quantify the effect of the mutation and circuit characteristics, algorithms and their classifications, and interactions (i.e., captured as independent variables) on the survivability of faulty benchmarks (the dependent variable), we define the metric \textit{Survival Rate} (\survivalRate). The \textit{survival rate} refers to the percentage of survived mutants (i.e., calculated based on undetected) obtained for a particular independent variable. This is the opposite of the typically used metric \textit{mutation score}, which is calculated based on the percentage of detected mutants. We decided to use the \survivalRate since we wanted to study the characteristics of mutants that are hard to detect; therefore, we chose a metric that focuses on mutants that are not detected. The metric is calculated by dividing the number of survived mutants by the total number of mutants, corresponding to each independent variable as:
\[\mathit{SR}_{\mathit{IV}} = \frac{\mathit{totalSurvivors}_{\mathit{IV}}} {\mathit{totalMutants}_{\mathit{IV}}}\]

\textit{IV} represents a set of independent variables corresponding to each mutant, algorithm, circuit characteristics, and interactions, e.g., \textit{Gate}. $\mathit{totalSurvivors}_{IV}$ represents the total number of survived mutants for a particular independent variable, e.g., for \textit{Gate} independent variable, an example is \textit{h} gate. $\mathit{totalMutants}_{\mathit{IV}}$ represents the total number of mutants for a particular independent variable (e.g., \textit{h} for \textit{Gate}).

In this study, we executed each mutant with its default program setting without introducing any specific input, i.e., all qubits initialized to 0. To determine the survival of a mutant, we check if a mutant is detected by the Wrong Output Oracle (WOO) and Output Probability Oracle (OPO) as was previously done in the literature~\parencite{Mendiluze2021,ourICST2021,quitoASE21tool}:
\begin{enumerate}
\item \textit{Wrong Output Oracle (WOO)}: The observed output does not match the expected output of the program, i.e., a new output is observed. As described in Section~\ref{subsubsec:algorithmcharac}, we have two broad categories of algorithms for \textit{Output Dominance}. For \textit{Output-dominant} algorithms, which are expected to produce one dominant output, if the dominant output produced by an algorithm does not match the expected dominant output, we consider it a detected mutant. For the rest of the algorithms (i.e., \textit{Diverse-output}), if any of the observed outputs do not match the expected outputs, we consider it as a detected mutant. 
\item \textit{Output Probability Oracle (OPO)}: The observed outputs match with the expected ones; however, the observed probabilities are significantly different than the expected ones. To compare the expected probabilities with observed probabilities, OPO employs the Chi-square test. We chose a significance level of 0.01, i.e., if the p-value is less than 0.01, we conclude that a mutant is detected. Note that, since in \textit{Output-dominant} algorithms we only care about the dominant output, i.e., the output with the highest probability, we only consider this oracle for \textit{Diverse-Output} algorithms. In \textit{Output-dominant} algorithms, if the observed dominant output remains the same as the expected dominant output, even if the probability of the output has changed, the mutant will not be considered detected.
\end{enumerate}

RQ3 studies the relations between circuit characteristics (e.g., \textit{\#qubits}) and \survivalRate. We study the correlation between an independent variable and \survivalRate with the Pearson correlation test~\parencite{Cleophas2018}. Pearson's correlation coefficient (\(r\)) measures the linear relationship between two variables, ranging from \(-1\) to \(1\). A value close to \(-1\) indicates a strong negative correlation, while a value close to \(1\) indicates a strong positive correlation. A value near \(0\) suggests little to no linear correlation. To interpret the strength of the correlation, we follow the widely accepted guidelines established by~\parencite{cohen1988statistical}:
\begin{inparaenum}[(i)]
\item \textbf{Negligible}: \( |r| < 0.10 \)
\item \textbf{Weak}: \( 0.10 \leq |r| < 0.30 \)
\item \textbf{Moderate}: \( 0.30 \leq |r| < 0.50 \)
\item \textbf{Strong}: \( |r| \geq 0.50 \).
\end{inparaenum}

\subsection{Subject Systems} \label{subsec:subjectsystems}
In our empirical study, we used a set of real quantum circuits provided by MQT Bench~\parencite{quetschlich2023mqtbench}. In total, MQT Bench offers more than 700,000 circuits with different configuration settings. We used the selection criteria provided by MQT Bench to obtain the circuits. The MQT Bench provides two types of circuits: Non-scalable circuits with fixed numbers of qubits and scalable ones, i.e., the same circuits implemented with different numbers of qubits. For scalable benchmarks, we configured the range of the number of qubits from 2 to 30. The maximum number of 30 was chosen because this is the maximum number of qubits we could simulate on a classical computer with the IBM simulator for most algorithms. Setting the maximum qubits to 30 for some circuits resulted in a complex circuit with many quantum gates that became infeasible to execute on quantum simulators. Consequently, we reduced the number of qubits for such algorithms according to this practical constraint. Since the study was performed in the Qiskit simulator, we selected the Qiskit compiler option at the target-independent level.
The obtained benchmark consists of 382 circuits (21 non-scalable and 361 scalable ones), which are implemented in Open Quantum Assembly Language (QASM) V.2~\parencite{cross2017open} as the original benchmarks. We further automatically translated them to Qiskit to be compatible with Muskit, which currently can only generate mutants for Qiskit. Table~\ref{tab:ProgramsCharacteristics} shows the characteristics of the original benchmarks for each quantum algorithm.
\begin{table}[!t]
\centering
\caption{Characteristics of the original benchmarks (i.e., quantum algorithm). ID represents the unique identifier of a quantum algorithm, whereas the bottom of the table shows which ID maps to which algorithm. An interested reader can check~\parencite{quetschlich2023mqtbench} for more algorithm details. For each algorithm, \#Q, \#G, \#M, D, \#SG, \#MQG, and \#EQ denote the minimum and maximum number of qubits, gates, measurements, depth, single qubit gates, multi-qubit gates, and entangled qubits. }
\label{tab:ProgramsCharacteristics}
\resizebox{\linewidth}{!}{\begin{tabular}{cccccccc|cccccccc}
\toprule
{\textbf{ID}} & \textbf{\#Q} & \textbf{\#G} & \textbf{\#M} & \textbf{D} & \textbf{\#SG} & \textbf{\#MQG} & \textbf{\#EQ} & {\textbf{ID}} & \textbf{\#Q} & \textbf{\#G} & \textbf{\#M} & \textbf{D} & \textbf{\#SG} & \textbf{\#MQG} & \textbf{\#EQ} \\
\midrule
QP1 & 2-20 & 8-287 & 2-20 & 6-114 & 5-77 & 3-210 & 0 & QP15 & 2-25 & 5-338 & 2-25 & 5-51 & 2-25 & 3-313 & 0 \\
QP2 & 2-30 & 5-89 & 1-29 & 4-32 & 3-59 & 2-30 & 0-22 & QP16 & 2-25 & 7-363 & 2-25 & 7-53 & 3-26 & 4-337 & 2-25 \\
QP3 & 2-30 & 3-31 & 2-30 & 3-31 & 1 & 2-30 & 2-30 & QP17 & 2-13 & 6-105 & 2-13 & 4-26 & 4-26 & 2-79 & 0 \\
QP4 & 3-30 & 7-61 & 3-30 & 5-10 & 3-30 & 4-31 & 0 & QP18 & 2-25 & 5-358 & 1-24 & 4-66 & 3-49 & 2-309 & 0 \\
QP5 & 14 & 225 & 14 & 43 & 42 & 183 & 0 & QP19 & 2-25 & 5-362 & 1-24 & 4-74 & 3-49 & 2-313 & 0 \\
QP6 & 12 & 169 & 12 & 37 & 36 & 133 & 0 & QP20 & 3-4 & 26-191 & 3-4 & 20-170 & 13-94 & 13-97 & 0-2 \\
QP7 & 4 & 25 & 4 & 13 & 12 & 13 & 0 & QP21 & 3-13 & 26-266 & 3-13 & 20-230 & 13-43 & 13-223 & 2-2 \\
QP8 & 2-6 & 3-451 & 2-6 & 2-402 & 2-98 & 1-353 & 0-5 & QP22 & 2-17 & 12-477 & 2-17 & 8-70 & 8-68 & 4-409 & 0 \\
QP9 & 2-7 & 3-311 & 2-7 & 2-271 & 2-46 & 1-265 & 0-5 & QP23 & 2-12 & 12-82 & 2-12 & 8-20 & 8-48 & 4-34 & 0 \\
QP10 & 3-11 & 23-211 & 6-22 & 15-47 & 12-44 & 11-167 & 0 & QP24 & 2-16 & 12-425 & 2-16 & 8-66 & 8-64 & 4-361 & 0 \\
QP11 & 3-17 & 22-477 & 3-17 & 14-70 & 12-68 & 10-409 & 0 & QP25 & 4-16 & 40-172 & 4-16 & 18-30 & 24-96 & 16-76 & 0 \\
QP12 & 5-15 & 43-384 & 5-15 & 36-343 & 22-180 & 21-204 & 0 & QP26 & 2-17 & 12-477 & 2-17 & 8-70 & 8-68 & 4-409 & 0 \\
QP13 & 5-15 & 43-402 & 5-15 & 36-347 & 22-198 & 21-204 & 0 & QP27 & 3-19 & 14-94 & 3-19 & 8-24 & 9-57 & 5-37 & 0 \\
QP14 & 3-15 & 17-77 & 6-30 & 10-12 & 9-45 & 8-32 & 0 & QP28 & 2-30 & 6-118 & 2-30 & 5-61 & 3-59 & 3-59 & 0 \\
\bottomrule
\end{tabular}}
\\
\justifying
\tiny*QP1: Amplitude Estimation (ae); QP2: Deutsch-Jozsa (dj); QP3: Greenberger-Horne-Zeilinger State (ghz); QP4: Graph State (graphstate); QP5: Ground State (groundstatelarge); QP6: Ground State (groundstatemedium); QP7: Ground State (groundstatesmall); QP8: Grover Search without Ancilla (grover-noancilla); QP9: Grover Search with Ancilla (grover-v-chain); QP10: Portfolio Optimization with QAOA (portfolioqaoa); QP11: Porfolio Optimization with VQE (portfoliovqe); QP12: Pricing Call Option (pricingcall); QP13: Pricing Put Option (pricingput); QP14: Quantum Approximate Optimization Algorithm (qaoa); QP15: Quantum Fourier Transform (qft); QP16: Quantun Fourier Transform Entangled (qftentangled); QP17: Quantum Generative Adversarial Networks (qgan); QP18: Quantum Phase Estimation Exact (qpeexact); QP19: Quantum Phase Estimation Inexact (qpeinexact); QP20: Quantum Walk without Ancilla (qwalk-noancilla); QP21: Quantum Walk with Ancilla (qwalk-v-chain); QP22: Real Amplitudes ansatz with Random Parameters (realamprandom); QP23: Routing Algorithm (routing); QP24: Efficient SU2 ansatz with Random Parameters (su2random); QP25: Travelling Salesman (tsp); QP26: Two Local ansatz with random parameters (twolocalrandom); QP27: Variational Quantum Eigensolver (vqe); QP28: W-State (wstate).

\end{table}

\subsection{Mutant Generation} \label{subsec:mutantGeneration}

We use the Muskit tool~\parencite{Mendiluze2021} to generate faulty benchmarks by applying \textit{Add}, \textit{Remove}, and \textit{Replace} mutation operators. To be comprehensive, for this empirical study, we applied all mutation operators combined with a total of 19 available gate types, on all possible positions, in each original quantum circuit for each quantum algorithm. The \textit{Add} operation is applied in all the possible positions using all supported gates, the \textit{Replace} operation replaces an existing gate with a new supported gate, and the \textit{Remove} operation will just remove an existing gate.

\begin{figure*}[!tb]
\includegraphics[width=\linewidth]{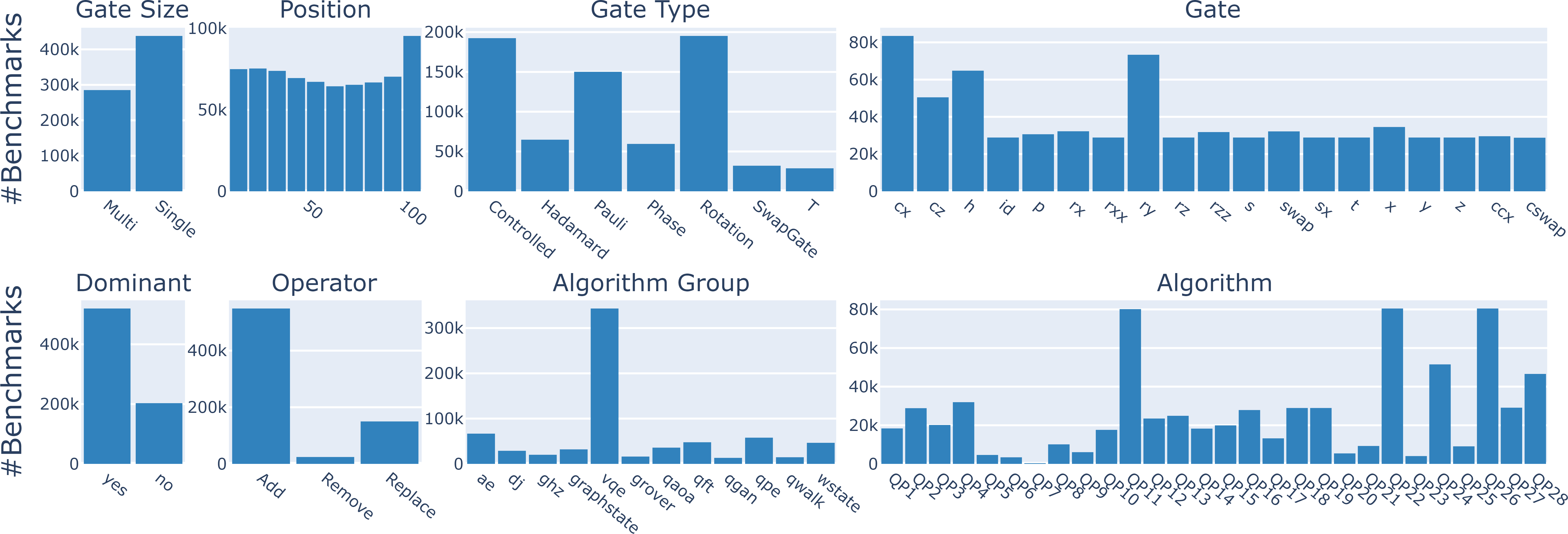}
\caption{Descriptive statistics of the generated benchmarks}
\label{fig:VariablesCountsBarPlot}
\end{figure*}

In the end, we obtained 723,079 faulty benchmarks, most of which were created with the \textit{Add} operator (75\%), followed by \textit{Replace} (20\%) and \textit{Remove} (3\%). Note that, as mentioned above, the \textit{Replace} and \textit{Remove} operators can only be applied on existing gates of the original quantum circuits, while the \textit{Add} operator can be applied anywhere in the circuit; therefore, compared with \textit{Add}, we obtained fewer numbers of faulty benchmarks created with them. We calculate the \survivalRate for each independent variable; note that, in this way, each operator is treated equally, as the number of its occurrences can not bias the results. Figure~\ref{fig:VariablesCountsBarPlot} presents the descriptive statistics of the generated benchmarks.

\subsection{Experimental Setup and Execution} \label{subsec:experimentalsetupexecution}
All the original and faulty benchmarks were executed under the same conditions using the same computational resources. The experiments run on a national high-performance cluster of servers, including 2x AMD Epyc 7601 processors, 2TB RAM, an AMD Vega20 GPU, and a high-speed 4TB NVMe drive. We performed a total of 100,000 shots for each circuit to deal with the inherent uncertainty in quantum computing. All the programs were executed using the Qiskit 0.43.1 version's Aer simulator to execute quantum circuits. To ensure consistency and reproducibility, we employed a fixed random seed, a key parameter of the Aer simulator for executing all quantum circuits, to deal with the inherent uncertainty of the quantum circuit execution.

\subsection{Threats to validity}\label{subsec:threatsToValidity}
A typical threat to the validity is about the generalization of our results. Note that we pick the real quantum algorithms from the most comprehensive circuit benchmark repository (i.e., MQT Bench). 
We generated more than 700K mutants from them, thus, making it the largest study on studying quantum mutation characteristics. Nonetheless, one could further enlarge the empirical evaluation with more circuits, which we will pursue in the future. 

Concerning the maximum number of qubits, we used a maximum of 30 qubit programs, and the results may be different for circuits with a higher number of qubits. However, it is impossible to execute large qubit circuits on classical computers. Moreover, one could also ask if our results are replicable on real quantum computers since we executed the circuits on the ideal simulator on classical computers. We argue that performing the study on the ideal simulator is important in our context since running experiments on real quantum computers would impact the results due to noise and could potentially lead to invalid conclusions. Nonetheless, we need a dedicated empirical evaluation on noisy quantum computers. 

We set the number of shots to 100K, which may be insufficient for large circuits producing all possible outputs. However, most of our algorithms produce dominant outputs; therefore, 100K shots are sufficient. For the OPO test oracle, we chose a significance level of 0.01. Using a lower threshold (e.g., 0.001) may lead to detecting more mutants, thereby decreasing \survivalRate. However, 0.01 is commonly used and is a reasonable choice~\parencite{DOMINGUEZJIMENEZ20111108,mogos2016quantum,CTQuantumQRS2021}. 

In the WOO oracle for the \textit{output-dominant} algorithms, we only checked whether the observed dominant output matches the expected one. This could impact survivability. However, there are not many research results on how to check the correctness of \textit{output-dominant} algorithms, thereby requiring more research on test oracles for such algorithms.

One of the main concerns in mutation analysis is the presence of equivalent and redundant mutants. In this study, we did not explicitly address the issue arising from such mutants, which could influence the conclusions drawn. Given the probabilistic nature of quantum programs and the difficulty in verifying similarity in quantum outputs, assessing the equivalence of mutants becomes particularly challenging. More research in this direction is needed to understand better and mitigate the impact of such mutants on mutation analysis, which is one of our future works.

When considering the abstraction level and how the mutations are introduced in classical mutation analysis, several pieces of evidence show the relationship between mutants and real faults~\parencite{couplingEffect}. However, in quantum mutation analysis, even though there are existing benchmarks for faults, linking syntactically introduced faults to developer faults remains challenging. Zhao et al., in Bugs4Q~\parencite{zhao2021bugs4q,Bugs4QJSS2023}, offer a bug classification that shows circuit-level faults as a viable approach. In the future, additional research will be conducted to keep the study up-to-date by incorporating new mutation operators into empirical evaluation corresponding to new real faults. 
\vspace{30pt}
\section{Results and Analysis}\label{sec:results}
We present the empirical evaluation results. For the 382 circuits corresponding to the 28 algorithms, we generated a total of 723079 faulty benchmarks. After executing them, we obtained the overall \survivalRate of 48\%, against the 51\% that were detected with test oracle WOO, and 1\% detected with test oracle OPO.

\subsection{Results for RQ1 -- Analyzing \survivalRate by Mutation Characteristics} \label{subsec:RQ1results}

\subsubsection{Results for RQ1.1 (individual characteristics)} 
\begin{figure*}[!htb]
\centering
\includegraphics[width=\linewidth]{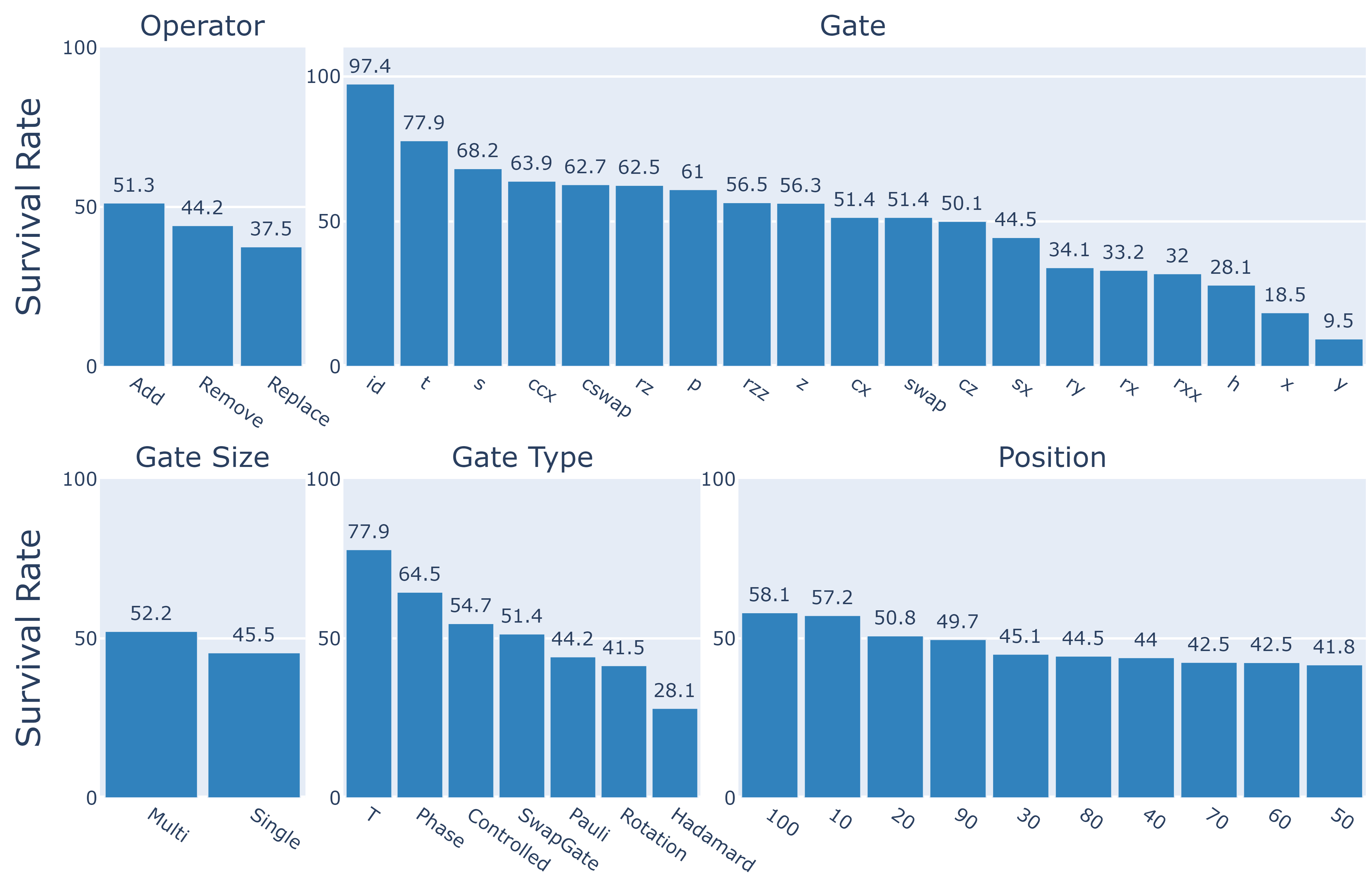}
\caption{Average \survivalRate of all faulty benchmarks in terms of each mutation characteristic -- RQ1.1}
\label{fig:MutantCharacteristicsSR}
\end{figure*}

Figure~\ref{fig:MutantCharacteristicsSR} presents the \survivalRate's descriptive statistics in each characteristic across all the faulty benchmarks. 
For \textit{Operator}, we observe that mutation operator \textit{Add} achieved higher \survivalRate (see Figure~\ref{fig:MutantCharacteristicsSR}) than \textit{Remove} and \textit{Replace}.
This observation provides evidence that the \textit{Add} operator is more likely to generate faulty benchmarks with high chances of surviving, which is often wanted for assessing testing methods. 
For \textit{Position}, we can notice that manipulating faults at the beginning and end of a quantum circuit (100\%, 90\%, 10\%, and 20\%) achieved the top four \survivalRate (around 50\% and above) among the ten categories. 
As for \textit{Gate Size}, whether being added, deleted, and replaced gate via a mutation operator is a single-qubit gate or multi-qubit gate does not lead to a big difference in \survivalRate, i.e., 45.5\% and 52.2\%, respectively. 
Regarding characteristic \textit{Gate}, we note that gate \textit{Id} achieved the highest \survivalRate. This is expected as \textit{Id} is a single-qubit gate that does not change the qubit's state. It is often used for error correction, fault tolerance, or as a placeholder for maintaining the same circuit depth. However, surprisingly the \textit{id} gate did not achieve a 100\%. Our investigation found that the remaining mutants that were detected easily were for \textit{Graphstate} (QP4). This algorithm produces all possible outputs with certain probabilities. Given our chosen number of shots (i.e., 100,000), we could not ensure covering all outputs for its implementation with more than 16 qubits, and therefore, we obtained false positives. 

On the other hand, gates \textit{x}, \textit{y}, and \textit{h} achieved the lowest \survivalRate since these gates introduce big changes in circuit logic, i.e., \textit{h} introduces superposition, whereas \textit{x} flips the state of a qubit (i.e., $\ket{1}$ to $\ket{0}$ and vice versa) and \textit{y} in addition to state flip, also rotates the phase about the Y axis by $\pi$ radians. As a result, survival rates were the lowest.

When looking at \textit{Gate Type}, we notice that the \textit{T} gates, \textit{Phase} gates, and \textit{Controlled} gates achieved the top three \survivalRate. On the other hand, the \textit{Hadamard} gate achieved the lowest \survivalRate. This is because the Hadamard creates superposition, and manipulating a fault with Hadamard changes the logic of a quantum circuit, thereby detecting the fault more easily compared with the other gates.

\subsubsection{Results for RQ1.2 (pair-wise interactions)}
\begin{figure}[!tb]
\includegraphics[width=\linewidth]{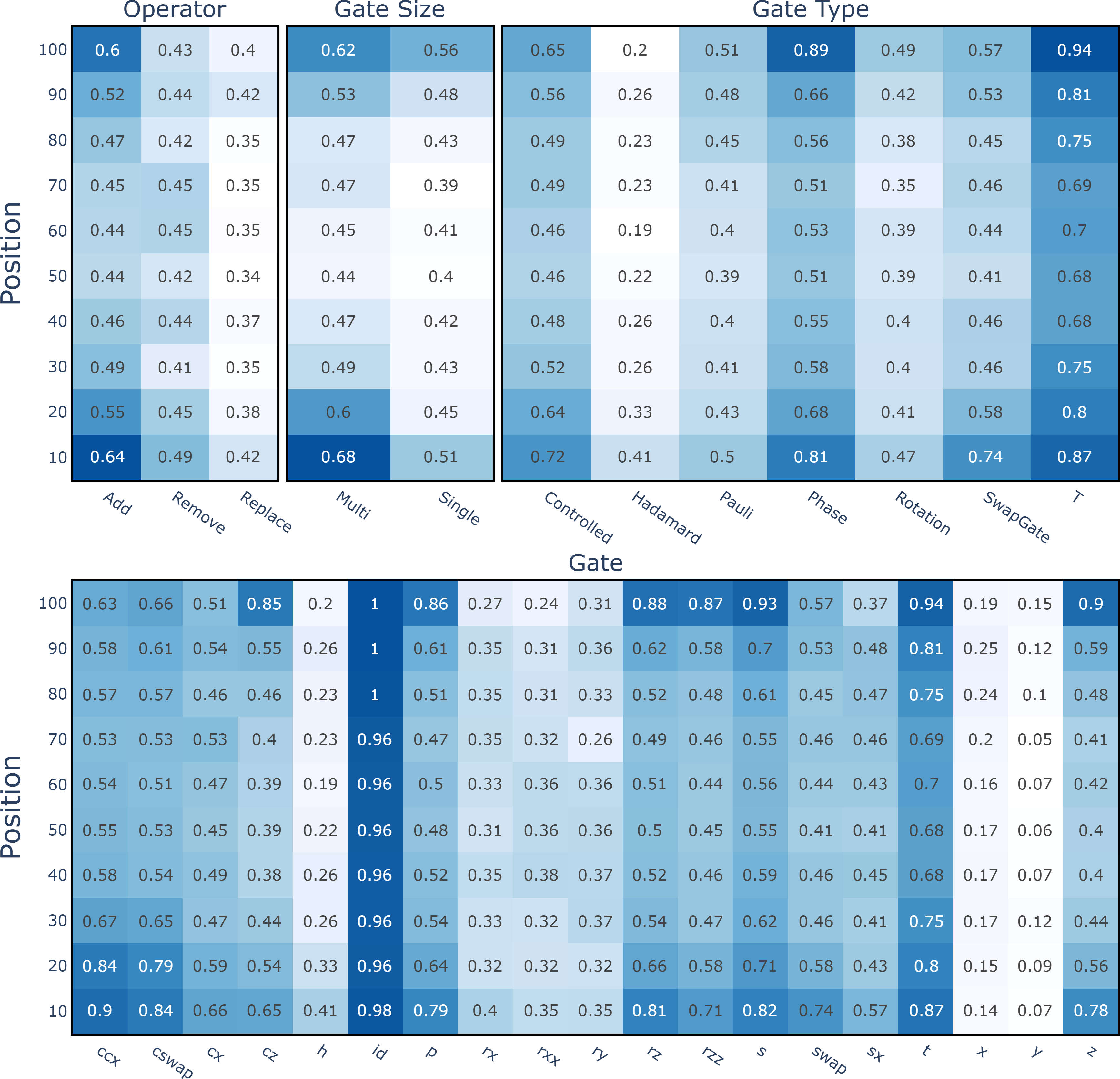}
\caption{Interaction effects between \textit{Position} and all other mutation characteristics -- RQ1.2. Each cell shows the \survivalRate corresponding to a specific interaction, with a darker (or lighter) blue indicating a higher (or lower) \survivalRate.}
\label{fig:HeatmapPositions}
\end{figure}

Figure~\ref{fig:HeatmapPositions} shows the effect of the interactions between \textit{Position} and each circuit characteristic (e.g., \textit{Gate}, \textit{Gate Size}).
Overall, most of the characteristics follow the same pattern mentioned above: faults applied at the beginning or end of a quantum circuit achieve a higher \survivalRate than the ones applied in the middle. Such a pattern can be observed clearly in \textit{Phase} gates or \textit{T} gates. However, not all the gates follow the same pattern: in the case of the \textit{Hadamard} gate, it starts with a higher \survivalRate at the beginning of the circuit and reaches the lowest \survivalRate at the end. In addition, the susceptibility of the different \textit{Gates} with respect to the \textit{Position} can be deduced; for instance, the interaction effect between \textit{Position} and gates \textit{x} and \textit{y} is very low, meaning that such gates are less susceptible to the \textit{Position} where they are applied.

\begin{figure}[!tb]
\includegraphics[width=\linewidth]{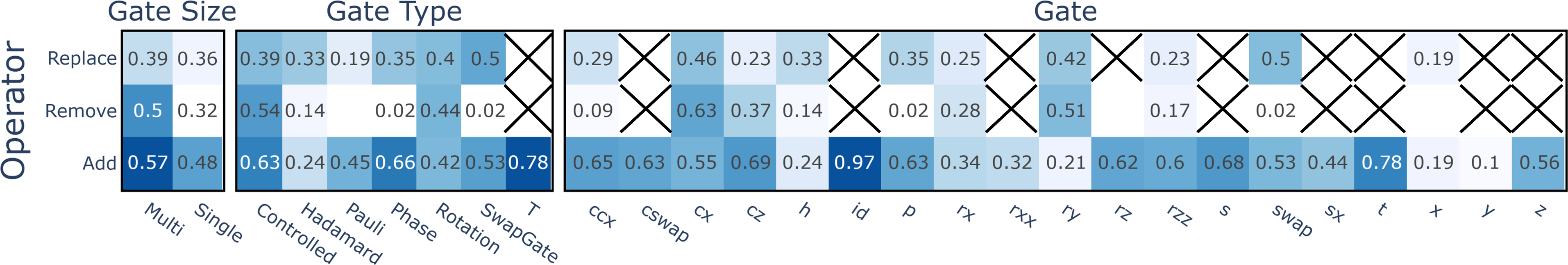}
\caption{Interaction effects between \textit{Operator} and all other mutation characteristics -- RQ1.2. Each cell shows the \survivalRate corresponding to a specific interaction. A darker (or lighter) blue indicates a higher (or lower) \survivalRate; a white empty cell denotes an absolute zero \survivalRate; a cell with zero in it denotes a very-near-zero positive number; a cell with X tells that no benchmarks can be generated with the given combination.}
\label{fig:HeatmapOperator}
\end{figure}

Figure~\ref{fig:HeatmapOperator} presents the interaction effects of \textit{Operator} with all other characteristics.
Note that in some cases (denoted with $\times$ in cells), the combinations (e.g., removing a T gate, replacing an id gate) are impossible for given quantum circuits since the original circuits do not contain these gates. Therefore, no corresponding faulty benchmarks were generated. From the figure, we can observe that, in most cases, the \textit{Add} operator is prominent in leading to high \survivalRate, as we have also discussed in RQ1.1. However, there are some exceptions. For instance, adding, removing, and replacing the rotation gates do not differ significantly.


\subsubsection{Results for RQ1.3 (interactions among all characteristics)}
To illustrate the categories' interaction effects, in Table~\ref{tab:TripleCombinationsAlg}, we report the top five cases that achieved the highest \survivalRate.
\begin{table}[!tb]
\centering
\small
\caption{Top 5 interactions of mutation characteristics \textit{Operator}, \textit{Gate} (or \textit{Gate Type}, \textit{Gate Size}) and \textit{Position} that achieved the highest \survivalRate, e.g., $Add\_id\_80.0\{1.0\}$: adding an \textit{id} gate at position 80\% achieved 100\% \survivalRate -- RQ1.3.}

\label{tab:TripleCombinationsAlg}
\resizebox{\linewidth}{!}{
\begin{tabular}{cccccc}
\toprule
\textbf{Combination} & \textbf{Top 1} & \textbf{Top 2} & \textbf{Top 3} & \textbf{Top 4} & \textbf{Top 5} \\ \midrule
Operator\_Gate\_Position & Add\_id\_80.0\{1.0\} & Add\_id\_90.0\{1.0\} & Add\_id\_100.0\{1.0\} & Add\_id\_70.0\{0.98\} & Add\_id\_10.0\{0.97\} \\
Operator\_Gate Type\_Position & Add\_T\_100.0\{0.92\} & Add\_T\_90.0\{0.89\} & Add\_Phase\_100.0\{0.86\} & Add\_T\_80.0\{0.83\} & Add\_T\_10.0\{0.83\} \\
Operator\_Gate Size\_Position & Remove\_Multi\_100.0\{0.71\} & Remove\_Multi\_90.0\{0.71\} & Add\_Multi\_10.0\{0.69\} & Remove\_Single\_100.0\{0.67\} & Add\_Multi\_100.0\{0.64\} \\
\bottomrule
\end{tabular}}
\end{table}
However, all the results are available in the online repository~\parencite{GitHubRepository}. When looking at the first row of the table, one can observe that adding \textit{id} gate at positions 80\%, 90\%, 100\%, 70\%, and 10\% achieved the top five \survivalRate (ranging from 100\% to 97\%). Regarding the effect of the interactions among \textit{Operator}, \textit{Gate Type} and \textit{Position}, we can observe, from the second row of the table, that adding a \textit{Phase} or \textit{T} gate at the end or beginning positions achieved the top five \survivalRate. Regarding the interaction effects of the combination of \textit{Operator}, \textit{Gate Size} and \textit{Position}, from the results of our study, we recommend adding a multi-gate at position 100\% to generate faulty benchmarks that are most challenging, i.e., the highest \survivalRate.

\begin{tcolorbox}[colback=blue!5!white,colframe=white,breakable]

\textbf{Concluding Remarks for RQ1:} Applying operator \textit{Add} led to slightly higher \survivalRate than \textit{Remove} and \textit{Replace}. Introducing faults at the beginning or end of a quantum circuit has a higher chance of generating faulty benchmarks that can survive testing. Gates \textit{y}, \textit{x} and \textit{h} achieved the lowest \survivalRate, while \textit{T} and \textit{Phase} achieved the highest \survivalRate. 
\end{tcolorbox}

\subsection{Results for RQ2 -- Analyzing \survivalRate by Algorithms and their Categorization}\label{subsec:RQ2results}

\subsubsection{Results for RQ2.1 (Output Dominance)} 
\begin{figure*}[!tb]
\centering
\includegraphics[width=1\textwidth]{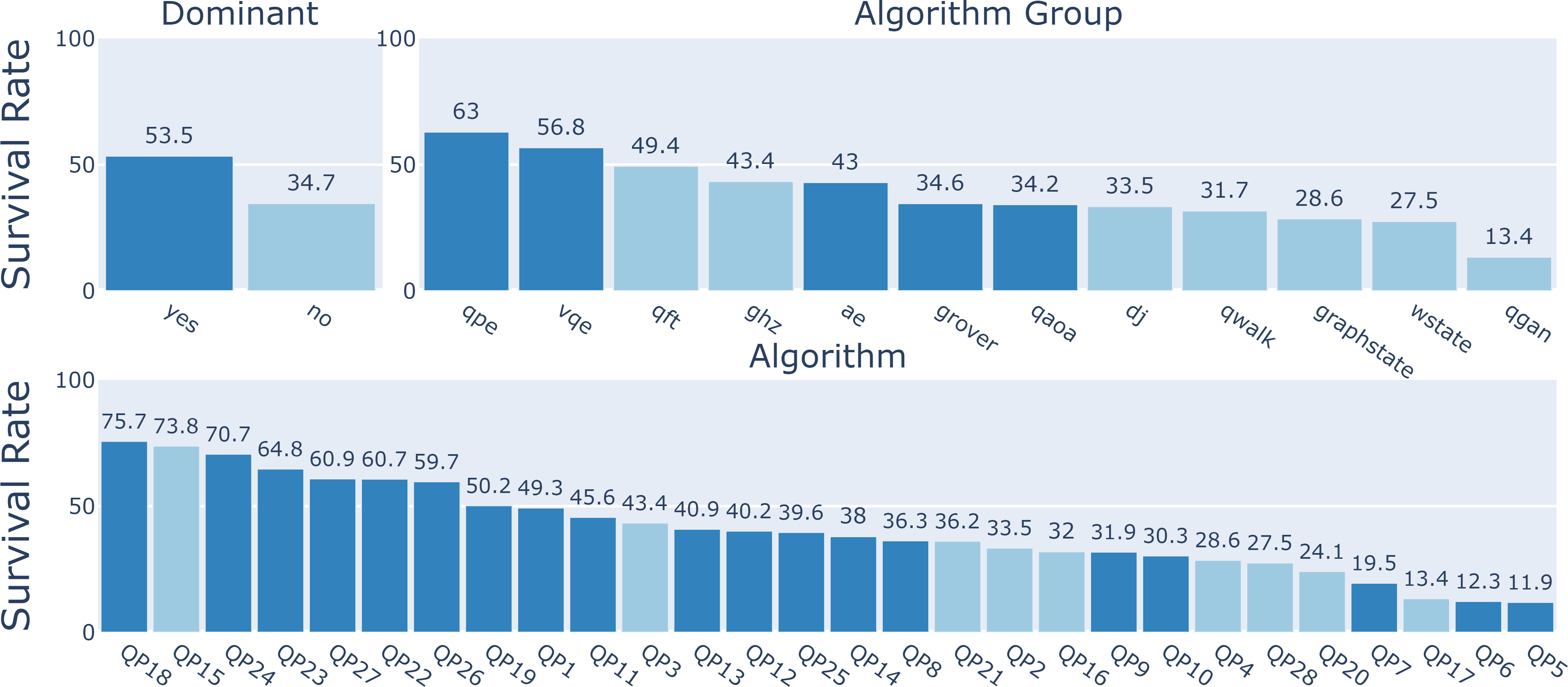}
\caption{Average \survivalRate of all faulty benchmarks regarding algorithms -- RQ2.
Dark and light blues differentiate output-dominant and diverse-output algorithms.}
\label{fig:algorithmRanking}
\end{figure*}

When comparing algorithms regarding the type of output (i.e., \textit{output-dominant} algorithms and \textit{diverse-output} algorithms), we observe that the \textit{output-dominant} algorithms have relatively higher \survivalRate (i.e., 53.5\%) than the others (i.e., 34.7\%), as shown in Figure~\ref{fig:algorithmRanking}.
This tells that faulty benchmarks of the \textit{output-dominant} algorithms are easier to survive.
Recall from Section~\ref{subsec:metrics} that, to check whether a faulty benchmark of an \textit{output-dominant} algorithm has survived, we assess the WOO oracle, i.e., checking whether the observed dominant output is the same as the expected. Therefore, if a faulty benchmark of \textit{output-dominant} algorithm does not change the expected dominant output, the assessment of WOO remains the same, meaning that the mutation cannot affect the dominant output, i.e., the benchmark survived.



\subsubsection{Results for RQ2.2 (Algorithm Group)} When looking into the 12 algorithm groups (see Figure~\ref{fig:algorithmRanking}), the faulty benchmarks generated for \textit{qpe} and \textit{vqe} achieved the highest \survivalRate: 63\% and 56.8\%, respectively, indicating that they are more tolerant to seeded faults. On the lowest side, \textit{qgan}, \textit{wstate}, and \textit{graphstate} obtained the lowest \survivalRate: 13.4\%, 27.5\%, and 28.6\%, respectively, indicating that the faulty benchmarks generated from these groups are difficult to survive. As already observed in RQ2.1, in general, \textit{output-dominant} algorithms tend to have higher \survivalRate. However, \textit{qft} and \textit{ghz} algorithms are exceptions, as they achieved higher \survivalRate than three other \textit{output-dominant} algorithms. For \textit{ghz}, one plausible explanation is that the \textit{ghz} algorithms entangle all qubits in a circuit, and once they are entangled, any mutation operator applied to one qubit will affect the state of all others; when they are measured, they will be all in the same basis state (i.e., all 0 or all 1). This logic seems to reduce the chance of producing incorrect outcomes even when faults are seeded, leading to high \survivalRate. As for the \textit{qft} algorithms, they are typically implemented with \textit{CP} and \textit{H} gates. However, \textit{CP} is not one of the gates that the three mutation operators can manipulate, as Muskit currently does not support it. Therefore, no cases exist for removing or replacing \textit{CP} gates in the generated benchmarks for \textit{qft}. Furthermore, as we already observed in RQ1.1, applying \textit{Remove} and \textit{Replace} led to lower \survivalRate, as compared to \textit{Add}. Therefore, relying on adding gates probably led to the high \survivalRate. However, we need additional experiments to understand \textit{qft} and \textit{ghz} better.

\subsubsection{Results for RQ2.3 (individual algorithm, i.e., Algorithm)} For each algorithm, we observe a similar pattern as in RQ2.2, where \textit{output-dominant} algorithms tend to have higher \survivalRate. The exceptions are the two \textit{diverse-output} algorithms: QP15 - \textit{qft} (belonging to the \textit{qft} algorithm group) and QP3 - \textit{ghz} (the only algorithm in the \textit{ghz} algorithm group), for which generated faulty benchmarks exhibited higher \survivalRate than some of the output-dominant algorithms. Note that we have explained the possible reason in RQ2.2.

When looking at the lowest \survivalRate, QP5 - \textit{groundstatemedium} and QP6 --\textit{groundstatelarge} and QP17 - \textit{qgan} obtained the lowest \survivalRate: 11.9\%, 12.3\%, and 13.4\% respectively. Interestingly, QP5 and QP6, even belonging to the \textit{output-dominant} algorithm category, still have the lowest SRs. One possible reason is that these two algorithms have a larger number of qubits as compared with QP7 - \textit{groundstatesmall} (Table~\ref{tab:ProgramsCharacteristics}), which all share the same overall structure. This explains that QP5 and QP6 have more possible outputs than QP7; consequently, the dominant output of both is less prominent and, therefore, more sensitive to faults. As for \textit{qgan}, among the \textit{diverse-output} algorithms, it has the lowest \survivalRate. This might be because \textit{qgan} has two key components: generator and discriminator. Any changes to the generator may influence the discriminator's behavior and vice versa, leading to a change that is propagated incrementally. Hence, \textit{qgan} is more sensitive to changes and therefore obtains low \survivalRate.

\begin{tcolorbox}[colback=blue!5!white,colframe=white,breakable]
\textbf{Concluding Remarks for RQ2:} 
Typically, the \textit{dominant-output} algorithms have higher survival rates except for ground state algorithms. For \textit{diverse-output}, in general, we observed low survival rates except for \textit{ghz} and \textit{qft}.
\end{tcolorbox}

\subsection{Results for RQ3 -- Analyzing \survivalRate by Gate and circuit complexity}

\begin{figure*}[!htb]
\includegraphics[width=\linewidth]{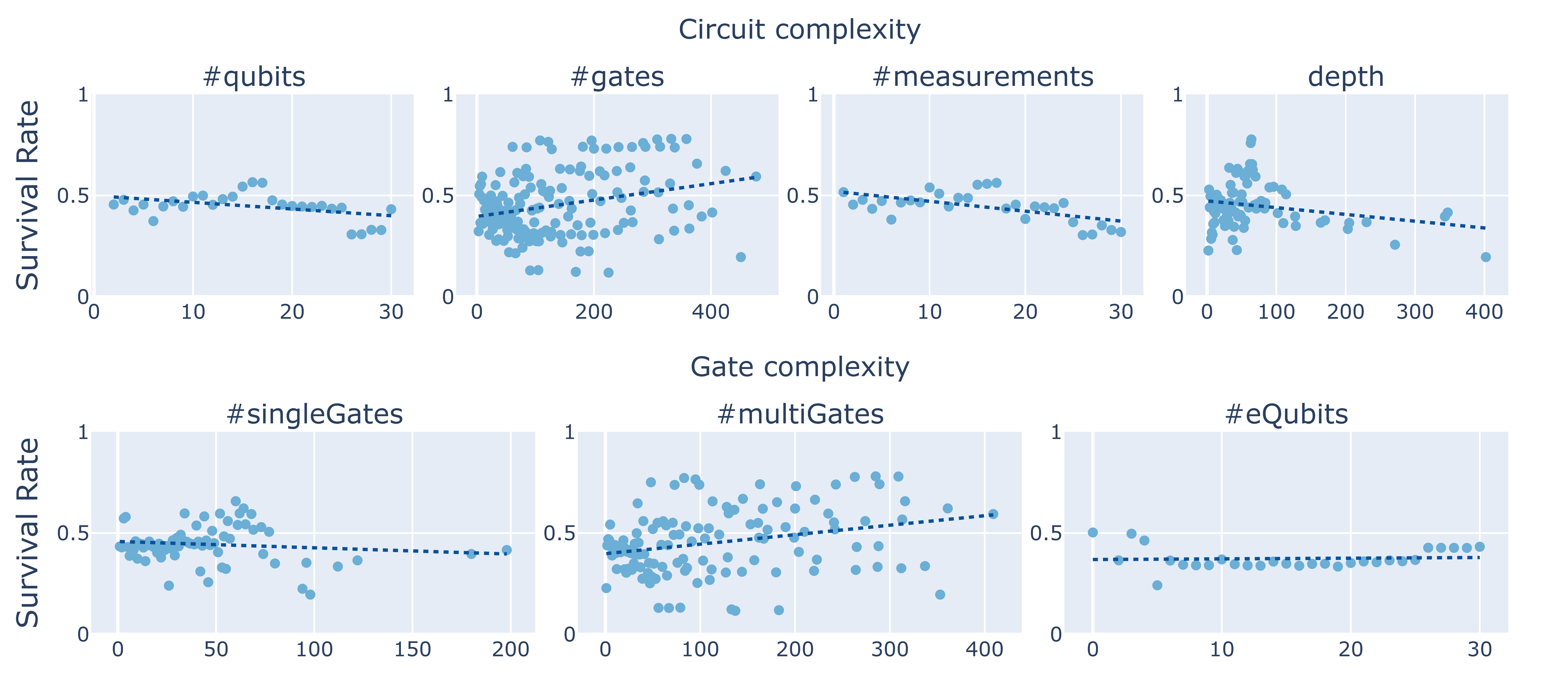}
\caption{Relationship between circuit and gate complexity variables with the \survivalRate.}
\label{fig:CorrelationPatterns}
\end{figure*}

We studied the correlation between independent variables (i.e., circuit and gate complexity variables) and the \survivalRate. Figure~\ref{fig:CorrelationPatterns} plots the correlation between each variable and its \survivalRate. The light blue dots refer to individual data points for each variable, while the dark blue dashed line represents a regression line indicating the trend.
The figure shows that \textit{\#gates} and \textit{\#multiGates} variables show a upward trend, while \textit{\#qubits}, \textit{\#measurements}, \textit{depth} and \textit{\#singleGates} variables show a downward trend. This suggests a potential positive or negative correlation, respectively. However, the data points around the regression lines suggest high variability, indicating that these relationships may not be strong.

We further applied the Pearson correlation test to assess whether these relationships are statistically significant. Table~\ref{tab:correlation_results} shows the results of the test.
\begin{table}[!htb]
\centering
\caption{Correlation tests results for circuit and gate complexity variables.}
\label{tab:correlation_results}
\begin{tabular}{llccc}
\toprule
\multicolumn{2}{c}{Variable} & Correlation & p-value & strength \\ \midrule
\multirow{4}{*}{\makecell{Circuit \\ complexity}} & \textit{\#qubits} & -0.007 & $<$ 0.05 & Negligible \\
&\textit{\#gates} & \textbf{0.13} & $<$ 0.05 & \textbf{Weak} \\
&\textit{\#measurements} & -0.033 & $<$ 0.05 & Negligible \\
&\textit{depth} & -0.032 & $<$ 0.05 & Negligible \\
\multirow{3}{*}{\makecell{Gate \\ complexity}} & \textit{\#singleGates} & 0.014 & $<$ 0.05 & Negligible\\
&\textit{\#multiGates} & \textbf{0.15} & $<$ 0.05 & \textbf{Weak}\\
&\textit{\#eQubits} & -0.082 & $<$ 0.05 & Negligible \\
\bottomrule
\end{tabular}
\end{table}
Results show that all correlation coefficients are between -0.082 and 0.15 for all circuit characteristics, indicating only negligible or weak correlation. This is expected as these characteristics, though they are important metrics for measuring the complexity of circuits, are static properties of the circuit and do not capture its behavior during execution. Since quantum circuits exhibit highly dynamic properties, such as state evolution and entanglement propagation, static metrics alone seem insufficient to predict the \survivalRate. Similar observations have been noted in prior studies, where circuit complexity measures were found to be weak indicators of execution behavior and computational efficiency~\parencite{cicero2024simulation}.

However, even though there is no strong correlation overall, we can observe that the most relevant variables are the number of multi-qubit gates (i.e., \textit{\#multiGates}) and the total number of gates (i.e., \textit{\#gates}) in the circuit. These variables appear to have a weak correlation with the survivability of the mutants, compared to the other variables having a negligible correlation, meaning that as the number of multi-qubit gates or the total number of gates in the original circuit increases, the \survivalRate of the mutant increases to some extent.


\begin{tcolorbox}[colback=blue!5!white,colframe=white,breakable]
\textbf{Concluding Remarks for RQ3:} 
No significant correlations can be observed between circuit characteristics and the survivability of faulty benchmarks. 
\end{tcolorbox}

\subsection{Results for RQ4 -- Interactions between Algorithm and Mutation Characteristics} \label{subsec:Rq4results}
\begin{figure*}[!tb]
\centering
\includegraphics[width=0.86\linewidth]{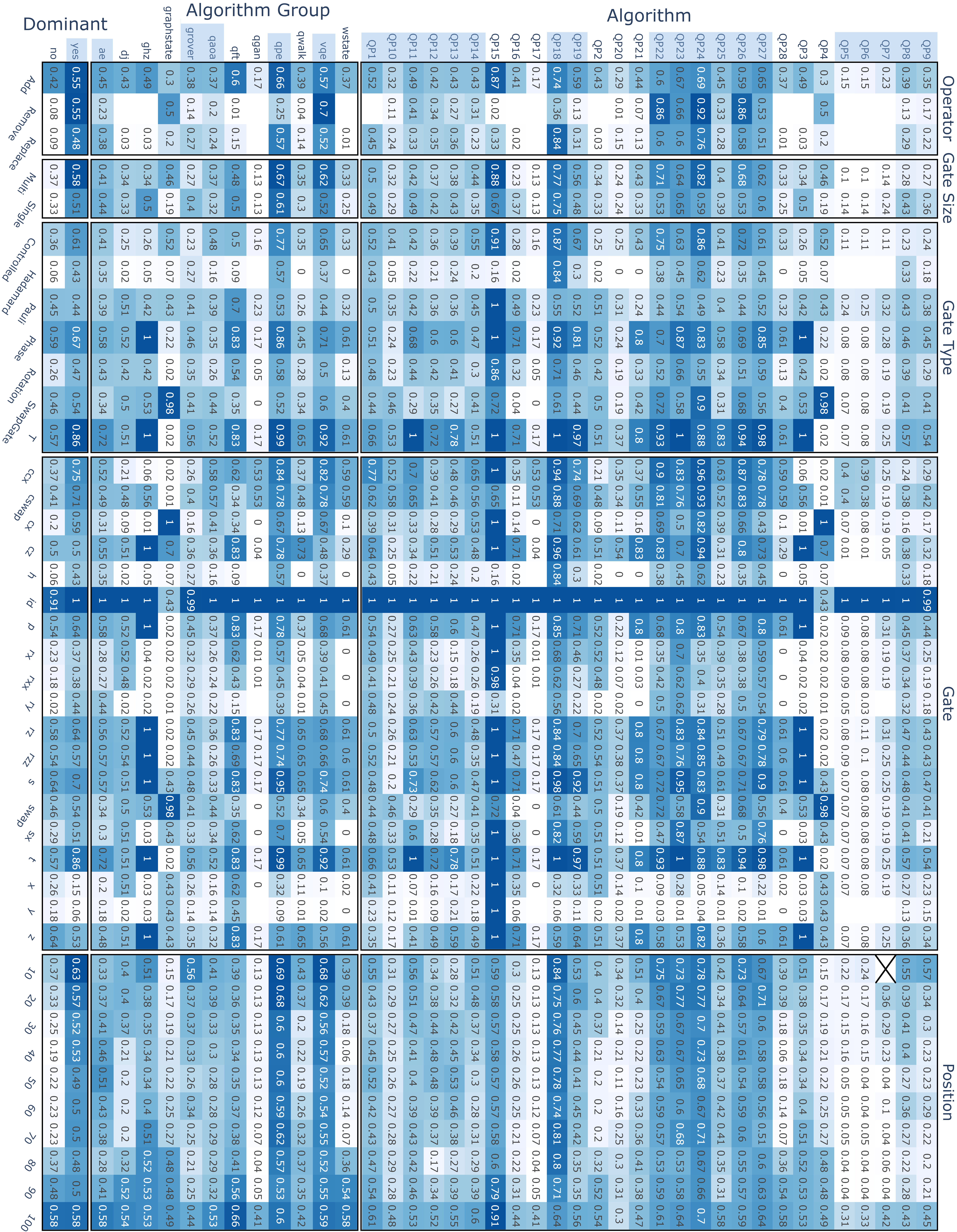}\\
\justifying
\tiny*QP1: ae; QP2: dj; QP3: ghz; QP4: graphstate; QP5: groundstatelarge; QP6: groundstatemedium; QP7: groundstatesmall; QP8: grover-noancilla; QP9: grover-v-chain; QP10: portfolioqaoa; QP11: portfoliovqe; QP12: pricingcall; QP13: pricingput; QP14: qaoa; QP15: qft; QP16: qftentangled; QP17: qgan; QP18: qpeexact; QP19: qpeinexact; QP20: qwalk-noancilla; QP21: qwalk-v-chain; QP22: realamprandom; QP23: routing; QP24: su2random; QP25: tsp; QP26: twolocalrandom; QP27: vqe; QP28: wstate \\
\caption{Interaction effects between \textit{Algorithm} (or \textit{Algorithm Group}, \textit{Dominant}) and all mutation characteristics -- RQ4. A darker (or lighter) blue indicates a higher (or lower) \survivalRate; a white empty cell denotes an absolute zero \survivalRate; a cell with zero in it denotes a very-near-zero positive number; a cell with X tells that no benchmarks can be generated with the given combination.}
\label{fig:HeatmapRQ4}
\end{figure*}

All results for RQ4 are presented in Figure~\ref{fig:HeatmapRQ4}.

\subsubsection{Results for RQ4.1 -- Interactions with Algorithm} When looking at the interactions of each algorithm with each mutation characteristic, one can notice that for most algorithms, applying the \textit{Add} operator led to the highest \survivalRate, as we already observed in RQ1. However, for certain algorithms (e.g., QP26--\textit{twolocalrandom}, QP24--\textit{su2random}, QP22--\textit{realamprandom}, which all belong to \textit{vqe}), the \textit{Remove} operator led to the highest \survivalRate, implying that removing a gate from these algorithms has the least impact on their outcomes.

Regarding the interaction of \textit{Algorithm} and \textit{Gate Size}, we observe that, in 15 out of 28 cases, manipulating multi-qubit gates achieves higher \survivalRate than manipulating single-qubit gates, supporting the conclusions obtained in RQ1.1. For instance, for QP4--\textit{graphstate}, manipulating multi-qubit gates led to 46\% \survivalRate, which is notably higher than what manipulating single-qubit achieved: 19\%. 
When looking at the interaction of \textit{Algorithm} and \textit{Gate Type}, manipulating a \textit{Hadamard} gate led to the least \survivalRate for most algorithms. However, there are a few exceptions. For instance, for QP18 - \textit{qpeexact}, manipulating a \textit{Hadamard} gate achieved 84\% \survivalRate, i.e., higher than manipulating a \textit{Pauli}, \textit{Rotation}, or \textit{Swap} gate. One plausible explanation is that the algorithm only has one dominant output, and therefore introducing a fault by manipulating a \textit{Hadamard} gate has a chance of changing the output probabilities but not necessarily altering the dominant output. 
Manipulating a \textit{T} gate led to the highest \survivalRate with a few exceptions, such as QP4 - \textit{graphstate}, for which manipulating a \textit{Swap} gate achieved the highest \survivalRate (i.e., 98\%). This is because this algorithm produces all possible outputs, so swapping the qubits does not have much effect, as the same state will be reached in another shot.

Regarding specific gates, as discussed in RQ1, \textit{id} led to the 100\% \survivalRate, with two exceptions on QP9 - \textit{grover-v-chain} and QP4 - \textit{graphstate}. For QP9, after checking the data, we noted that only the two-qubit circuit was affected by \textit{id} because switching the dominant output from one to the other is easy due to similar probabilities. As for QP4 - \textit{graphstate}, as discussed in RQ1.1, it produces all possible outputs, and with the given number of shots, we could not cover all possible outputs.

When comparing Pauli gates (\textit{X}, \textit{Y}, and \textit{Z}), \textit{Z} achieved the highest \survivalRate for most cases. This might be because \textit{Z} does not change the probabilities of outputs. For QP5 - \textit{groundstatelarge} and QP6 - \textit{groundstatemedium}, the differences in \survivalRate across the different gates excluding \textit{id} are very small. We note that for QP15 - \textit{qft}, 14 out of 19 gates achieved 100\% \survivalRate, implying that the faulty benchmarks for \textit{qft} are very difficult to detect, as already discussed in RQ2.2.

When checking \textit{Position}, we note that introducing faults to the beginning and end of the circuits of most algorithms led to high \survivalRate with some exceptions (e.g., QP17, QP18, QP4). For instance, for QP4 - \textit{graphstate}, the \survivalRate at position 10\% is only 0.15, implying that introducing a fault at the beginning of the circuit of the \textit{graphstate} algorithm is possible to change its behavior.

\subsubsection{Results for RQ4.2 (Interaction with Algorithm Group)} 
Figure~\ref{fig:HeatmapRQ4} shows that the \textit{vqe} and \textit{graphstate} algorithms achieved the highest \survivalRate when the \textit{Remove} operator is applied. For \textit{vqe}, removing a gate only changes the probability of a dominant output. In \textit{vqe}, we only care about the correct dominant output, and as long as the correct output remains dominant, it is considered survived.

Regarding \textit{Gate Size}, we only observe a difference in terms of \survivalRate between manipulating single-qubit gates and multi-qubit gates for \textit{grover} (40\% vs 27\%) and \textit{graphstate} (19\% vs 46\%). As we saw in Section~\ref{subsec:RQ1results}, multi-qubit, overall, had a higher \survivalRate than single-qubit, with which the behavior of \textit{graphstate} conforms. However, for \textit{grover}, it is reversed. A possible reason is that \textit{grover} has many multi-qubit gates (Table~\ref{tab:ProgramsCharacteristics}); therefore, removing or replacing multi-qubit gates has a high chance of changing the logic of the circuits, which, hence, led to lower \survivalRate than manipulating single-qubit gates.

Manipulating \textit{Hadamard} (resp. \textit{T}) achieved the lowest (resp. highest) \survivalRate for most algorithm types, as also discussed in RQ1. We further observe that \survivalRate of \textit{vqe} and \textit{qpe} are higher than those of other algorithms across the different gate types (also discussed in RQ2.2). 

For each individual gate, we observe differences in the \survivalRate among the same gate types. For instance, for the \textit{ghz} algorithms, manipulating \textit{cz} has a much higher chance of making the faulty \textit{ghz} benchmarks survive when compared with \textit{cx}, i.e., 100\% vs 1\% \survivalRate. This is reasonable because a \textit{cz} gate induces a phase flip, which might not affect the final measurement results, but a \textit{cx} gate introduces both bit and phase flips. When comparing the three Pauli gates (\textit{x}, \textit{y}, and \textit{z}), we observe that manipulating a \textit{z} gate led to much higher \survivalRate across the algorithm groups. Similarly, this might be because \textit{z} gates do not change the probabilities of outputs.

Regarding \textit{Position}, the \textit{vqe} and \textit{qpe} algorithms are less sensitive to the positions where the faults are seeded. This is because initializing the \textit{vqe} and \textit{qpe} algorithms is not only about applying the Hadamard gate to all qubits. Specifically, a parameterized trial state should be created for a \textit{vqe} algorithm as part of the initialization, and an eigenstate should be prepared for \textit{qpe}. Therefore, they both do not show the pattern we observed about \textit{Position}: seeding a fault at the beginning or end of a circuit is easy to survive (RQ1.2).

\subsubsection{Results for RQ4.3 -- Interaction with Output Dominance} 
As expected, results for the \textit{output-dominant} algorithms and \textit{diverse-output} algorithms are quite different. First, for \textit{output-dominant} algorithms (e.g., \textit{qpe}), our results do not reveal much difference in terms of the mutation operators (55\% vs 55\% vs 48\% for \textit{Add}, \textit{Remove} and \textit{Replace}, respectively). Also, the \survivalRate of these algorithms' faulty benchmarks is generally higher than the \textit{diverse-output} algorithms. This is reasonable because \textit{output-dominant} algorithms are generally more robust or fault-tolerant of faults, as their working mechanisms amplify the probability of the correct answer (i.e., dominant output). Even with small errors, the correct answer can still be the most likely outcome. Due to this reason, we think manipulating \textit{Hadamard} gates in the \textit{output-dominant} algorithms led to comparable \survivalRate as manipulating \textit{Pauli} and \textit{Rotation} gates, and they are not sensitive to where a fault is seeded. This is, however, not the case for \textit{diverse-output} algorithms.

\begin{tcolorbox}[colback=blue!5!white,colframe=white,breakable]
\textbf{Concluding Remarks for RQ4:} 
The interactions of \textit{Algorithm}, \textit{Algorithm Group}, and \textit{Output Dominance} with the mutation characteristics conform to what we observed in RQ1 and RQ2 with a few exceptions, mostly caused by individual algorithms' unique characteristics. We also observed the importance of distinguishing \textit{output-dominant} and \textit{diverse-output} algorithms in terms of where to seed a fault and which mutation operator to apply. 
\end{tcolorbox}

\section{Discussion and Recommendations} \label{sec:discussion}
This section explores the implications of our findings to provide a guide for practitioners and researchers. We discuss the availability of the dataset and recommendation tool in Section~\ref{subsec:availability} and their implications for researchers and practitioners in Sections~\ref{subsec:implicationsPra} and~\ref{subsec:implicationsRes} respectively. We also highlight the evolution of the findings in Section~\ref{subsec:evolution} and acknowledge the limitations of our output assessment method in Section~\ref{subsec:assessmentLim}.


\subsection{Availability of Dataset and Recommendation Tool}\label{subsec:availability} The dataset from our empirical evaluation is available for anyone interested in using it~\parencite{GitHubRepository}.\footnote{Original and faulty benchmarks can be downloaded from our GitHub repository~\parencite{GitHubRepository}.} The dataset contains all the mutants created during the study, along with their respective original circuits. The dataset includes a variety of quantum circuit benchmarks with diverse mutation characteristics and covers several quantum algorithms with different complexity levels and fault characteristics.

On top of the faulty benchmarks, we also built a software tool,\footnote{Note that the software tool is also available in the GitHub repository~\parencite{GitHubRepository}.} which can recommend faulty benchmarks to users based on selection criteria. This recommendation tool was built by studying all the possible interactions between \textit{Algorithm}, \textit{Algorithm Group}, or algorithms by \textit{Output Dominance} with all possible mutation characteristics (i.e., single, pair-wise, and three-ways). For instance, a user can specify an \textit{Algorithm}, \textit{Algorithm Group}, or algorithms by \textit{Output Dominance} together with desired \survivalRate and a maximum number of faulty benchmarks.

\subsection{Implications for Researchers} \label{subsec:implicationsRes} 
Given the limited availability of benchmarks with quantum faults, the provided dataset serves as a fault repository. Researchers can utilize this dataset to evaluate the effectiveness of their testing techniques in identifying faults across various quantum circuits. By comparing the performance of different techniques against a comprehensive set of faulty benchmarks, researchers gain a clearer understanding of how well their new testing techniques can detect and address potential faults, ultimately guiding the development of more reliable quantum software testing techniques.

The recommendation tool provides researchers with a structured approach for selecting mutants, enabling systematic studies into fault characteristics and their impact on quantum circuits. By applying these mutants to their circuits, researchers can analyze how different types of faults affect the circuit and identify potential weaknesses.

Furthermore, we generate new knowledge about relationships between the survival rate of faulty benchmarks and the characteristics of mutations, circuits, algorithms, and their interactions. This knowledge provides evidence about faulty benchmarks with which characteristics likely survive, based on which one can generate new faulty benchmarks for new quantum algorithms or algorithms not yet studied in our empirical study. With this knowledge, more advanced quantum mutation analysis techniques can be developed. For instance, an optimization approach, such as one based on search algorithms, could be utilized to minimize the number of selected faulty benchmarks.

\subsection{Implications for Practitioners}\label{subsec:implicationsPra} 
Mutation analysis is a costly approach, as it involves generating and evaluating a large number of mutants. The recommendation tool can help practitioners reduce this cost by selecting a diverse subset of mutants for their specific needs. For instance, in a use case involving a specific quantum algorithm, the tool can help the user by providing mutants with varying mutation characteristics and \survivalRate. Practitioners could assess whether their test suites can detect faults seeded in faulty benchmarks of varying survival rates. Moreover, one could also assess whether their test suites can identify faults seeded with specific mutation characteristics (e.g., the position of a fault).



\subsection{Evolution of the Findings}\label{subsec:evolution} 
Given that quantum software engineering is in its preliminary stage, one could argue that the mutants used in this study are too circuit-specific to be used in the future. Our empirical evaluation aims to explore the present state of quantum mutants and their characteristics. As the field advances, new quantum gates and circuit-based methodologies are likely to emerge, creating more complex quantum circuits. This evolution will require revisiting and updating our study to incorporate advancements, thereby generating fresh insights into quantum mutation analysis.

This study is designed to be adaptable. For instance, when a new mutation operator is introduced in Muskit due to a change in Qiskit (e.g., a new quantum gate is implemented), the new mutants corresponding to the new mutation operator will be generated and executed with Muskit. Next, the analysis will be rerun to obtain the updated findings. In this way, we keep our findings up to date as quantum computing frameworks continue to evolve.

\subsection{Output Assessment Method Limitations}\label{subsec:assessmentLim} 
This study employed one type of output assessment method based on output distributions. Selecting only one output assessment method could potentially bias our results toward this specific method. We chose this output assessment method since it has been commonly employed in quantum software testing literature~\parencite{quantumTestingRoadmapTOSEM2025}. However, other alternatives exist, such as output assessments based on state vectors and expectation values.

State vectors refer to the complete mathematical representation of a quantum state, providing full information about the system before measurement. The state vector approach requires full visibility of the quantum state to determine correctness, but obtaining the quantum state in real hardware is not feasible due to the measurement constraints inherent in quantum mechanics~\parencite{kalev2015quantum,tannu2019mitigating,malik2014direct}.

Expectation values in quantum circuits are statistical quantities derived from output distributions, requiring additional post-processing and measurements in different observables. Unlike raw output distributions, which provide probabilities of measurement outcomes, expectation values extract properties by averaging measurements. Since different observables may not share a common measurement basis, additional circuit executions with appropriate basis rotations are necessary, increasing computational demand~\parencite{kohda2022quantum}. Nonetheless, this remains an important area of study for the future in quantum mutation analysis.

Therefore, while our method may introduce some limitations, it is currently the most practical method for analyzing mutant survivability, especially in the context of real-world quantum hardware.

Future research could explore the impact of using multiple assessment methods on survivability results. This would help assess the applicability of our findings further and reduce any remaining bias introduced by the dependency on a single method. Nonetheless, our current approach provides a strong foundation for evaluating the survivability of mutants in quantum circuits.

\section{Conclusions and Future Work} \label{sec:conclusionAndRelated}
We presented an empirical study with more than 0.7 million quantum circuit mutants to study how various mutation characteristics, circuit characteristics, algorithm types, and their interactions relate to mutants being undetected (called survivability). Such a study helps to systematically design and generate faulty benchmarks for evaluating quantum software testing techniques' effectiveness from various aspects, e.g., the capability of detecting various types and complexity of faults. Based on the results, we provide actionable recommendations for researchers and practitioners to generate faulty benchmarks.

To expand the scope of our study, we plan to evaluate a variety of quantum benchmarks, including both larger circuits and different detection oracles. Additionally, we aim to evaluate these circuits on real quantum computers and noisy quantum simulators to understand better how different noise levels impact the survivability of mutants. This will allow us to assess the practical effectiveness of our findings in more realistic environments.

Moreover, we plan to extend our empirical analysis to include the detection of equivalent and redundant mutants. This will provide a more comprehensive understanding of how mutation operators affect quantum circuit behavior and performance, offering valuable insights for designing more robust quantum testing frameworks in both theoretical and practical contexts.

Additionally, we would like to investigate whether the characteristics used in our study (e.g., output dominance) could be reported into quantum computing frameworks such as Qiskit, so that developers can more easily use our findings to select the mutants to assess the test suites of their quantum programs.


\section*{Declarations}
\subsection*{Data Availability Statement}
All detailed experiment results, code, and data are available in the online repository~\parencite{GitHubRepository}.
\subsection*{Acknowledgements}
Eñaut Mendiluze Usandizaga is supported by Simula’s internal strategic project on quantum software engineering. Shaukat Ali is supported by the Qu-Test project (Project \#299827) funded by the Research Council of Norway, Oslo Metropolitan University's Quantum Hub, and Simula's internal strategic project on quantum software engineering. Paolo Arcaini is supported by Engineerable AI Techniques for Practical Applications of High-Quality Machine Learning-based Systems Project (Grant Number JPMJMI20B8), JST-Mirai.
\subsection*{Funding}
The research leading to these results received funding from: Simula Research Laboratory under Internal Strategic Project on Quantum Software Engineering, storbyuniversitetet under Quantum Hub Initiative, Norges Forskningsråd under Qu-Test project \#299827, JST-Mirai Program under Grant Agreement No JPMJMI20B8.
\subsection*{Conflicts of interests/Competing interests}
The authors have no competing interests to declare that are relevant to the content of this article.


\printbibliography

\end{document}